\documentclass[%
aip,jap,
sd,%
amsmath,amssymb,
reprint,%
]{revtex4-1}

\usepackage{graphicx}
\usepackage{dcolumn}
\usepackage{bm}
\usepackage{booktabs} 
\usepackage[final]{changes} 
\definechangesauthor[color=blue]{JAG} 
\definechangesauthor[color=red]{NOB}

\begin{document}
	
	\preprint{AIP/123-QED}
	\title[Critical Current Oscillations of Elliptical Josephson Junctions with Single-Domain Ferromagnetic Layers
	]{Critical Current Oscillations of Elliptical Josephson Junctions with Single-Domain Ferromagnetic Layers
	}
	
	\author{Joseph A. Glick}
	\author{Mazin A. Khasawneh}
	\author{Bethany M. Niedzielski}
	\author{Reza Loloee}
	\author{W. P. Pratt, Jr.}
	\author{Norman O. Birge}\email{birge@pa.msu.edu}
	\affiliation{Department of Physics and Astronomy, Michigan State University, East Lansing, MI 48824, USA}
	\author{E. C. Gingrich}
	\affiliation{Northrop Grumman Systems Corporation, Baltimore, MD 21240, USA}
	\author{P. G. Kotula}
	\author{N. Missert}
	\affiliation{Sandia National Laboratories, Albuquerque, NM 87185, USA}

	\date{\today}
	
	\begin{abstract}
		Josephson junctions containing ferromagnetic layers are of considerable interest for the development of practical cryogenic memory and superconducting qubits. Such junctions exhibit a ground-state phase shift of $\pi$ for certain ranges of ferromagnetic layer thickness. We present studies of Nb based micron-scale elliptically-shaped Josephson junctions containing ferromagnetic barriers of Ni$_{81}$Fe$_{19}$ or Ni$_{65}$Fe$_{15}$Co$_{20}$. By applying an external magnetic field, the critical current of the junctions are found to follow characteristic Fraunhofer patterns, and display sharp switching behavior suggestive of single-domain magnets. The high quality of the Fraunhofer patterns enables us to extract the maximum value of the critical current even when the peak is shifted significantly outside the range of the data due to the magnetic moment of the ferromagnetic layer. The maximum value of the critical current oscillates as a function of the ferromagnetic barrier thickness, indicating transitions in the phase difference across the junction between values of zero and $\pi$. We compare the data to previous work and to models of the 0-$\pi$ transitions based on existing theories.
		
	\end{abstract}
	
	\pacs{ }
	\keywords{Superconductivity, Josephson Junction, Ferromagnetism, Cryogenic Memory, Proximity Effect}
	\maketitle

	\section{\label{sec:level1}Introduction}
	
	Josephson junctions containing ferromagnetic materials (SFS junctions) have been under intense study for the past 15 years. The experimental breakthrough that triggered such intense interest was the demonstration that the ground-state phase difference across the junction can be either 0 or $\pi$, depending on the thickness of the ferromagnetic layer(s) in the junction~\cite{Ryazanov2001,Kontos2002}. That result has been confirmed by numerous groups since the initial discovery, using a wide variety of weak and strong ferromagnetic materials \cite{Blum2002,Sellier2003,Shelukhin2006,Weides2006,Robinson2006,Robinson2007,Bannykh2009}. There have been several proposals to use $\pi$-junctions as new components in either classical or quantum information processing circuits \cite{Terzioglu1998,Ioffe1999,Blatter2001, UstinovKaplunenko2003,Yamashita2005,Khabipov2010,Feofanov2010}.
	
	Our current work in this area focuses on the development of cryogenic random access memory \cite{Niedzielski2015,Gingrich2016}. Numerous ideas have been presented in the literature regarding how SFS junctions might be used as practical memory devices \cite{Bell2004,herr_patent2012,Larkin2012,Goldobin2013,Baek2014, Qader2014,herr_phasepatent2015}. The ferromagnetic (F) layer influences the properties of the junction both through the magnetic field and the exchange field it generates, and ideas have been presented using either of those mechanisms. In addition, either the critical current magnitude, $I_c$, or the ground-state phase difference across the junction, $\phi$, can be used as the physical quantity associated with information storage. Without trying to summarize the whole field, we can nonetheless make two general observations: 1) Proposals that rely on the internal magnetic field of the junction tend to become less viable as junction size decreases, since the relevant physical parameter defining the effect of the field is the magnetic flux, $\Phi$, threading the junction area. If the junction area is reduced to the point where $\Phi << \Phi_0$ = h/2e = 2.07$\times 10^{-15}$Tm$^2$, then the magnetic field has negligible effect on the junction properties. For that reason, we have chosen to emphasize the effect of the exchange field in our work. 2) Proposals that rely on the magnitude of $I_c$ invariably require that the SFS junction switch from the zero-voltage state into the voltage state when the memory is read. That switching process takes a time of order \added[id=JAG]{$\tau_{\mathrm{switch}} \approx \hbar/(eI_cR_N)$}, where $R_N$ is the junction resistance in the voltage state. For standard SFS junctions, $\tau_{\mathrm{switch}}$ is much too long to be useful for memory applications. \added[id=JAG]{One can shorten $\tau_{\mathrm{switch}}$ somewhat by increasing $R_N$ via the introduction of an insulating barrier to make an SIFS junction \cite{Weides2006,Larkin2012}. Suppression of $I_c$ by the insulating layer can be mitigated by adding a thin auxiliary nearly-superconducting (s) layer, to form an SIsFS junction with very large values of $I_cR_N$ \cite{Bakurskiy2013, Vernik2013, Ruppelt2015, Bakurskiy2017}.}
	
	\added[id=JAG]{An alternative scheme, proposed by workers at Northrop Grumman Corporation, envisions a memory cell consisting of a SQUID loop that contains a phase-controllable Josephson junction and two conventional SIS junctions with much smaller $I_c$'s~\cite{herr_phasepatent2015}.  The critical current of the SQUID loop is determined by the phase state of the controllable junction, either 0 or $\pi$, which correspond to the logical 0 or 1 of the memory.  During the read operation, only the SIS junctions may switch into the voltage state, providing a fast $\tau_{\mathrm{switch}}$, while the controllable junction remains in the superconducting state.  As a result, high $I_cR_N$ is not a critical requirement for the controllable junction in the Northrop Grumman design.  This is the memory design we are currently pursuing.}
	
	What we want, then, is a Josephson junction whose ground-state phase difference can be controllably switched between the 0 and $\pi$ states. It is advantageous that one or both of the nanomagnets in such a junction be single-domain, so that the magnetizations are uniform and magnetic switching is clean and reproducible. \added[id=JAG]{One method of accomplishing these goals} is to make a junction containing a ``spin valve", i.e. two F layers whose relative magnetization directions can be switched between parallel and antiparallel \cite{Bell2004,Qader2014,Baek2014}. We demonstrated such a controllable 0-$\pi$ junction recently, using Ni and NiFe as the two ferromagnetic materials~\cite{Gingrich2016}. Thin Ni films are magnetically ``hard", whereas NiFe is ``soft", hence it was possible to reverse the magnetization direction of the NiFe without changing that of the Ni. Our demonstration of reproducible 0-$\pi$ switching in spin-valve junctions was a success from the point of view of a demonstration, but the devices used in that experiment have some drawbacks for use in a practical cryogenic memory array. The biggest limitation of those devices was the poor magnetic properties of the Ni layers, which required being subjected to a large initialization field of 260 mT to set their magnetization direction before the start of the experiment. Applying such a large field is undesirable in a superconducting circuit; as it requires subsequent warming of the sample to just above the critical temperature of the Nb electrodes to remove trapped flux. Because of that limitation, we have been searching for a material to replace Ni as the hard magnetic layer in our spin-valve junctions. NiFeCo is a promising candidate: like NiFe, it has a small magneto-crystalline anisotropy whose direction can be set by depositing the material in the presence of a magnetic field, but its anisotropy is somewhat larger than that of NiFe. Using NiFeCo as the hard layer should allow us to use initialization fields of only a few tens of mT.
	
	Before using a new magnetic material in a spin-valve junction, one would like to know how it behaves by itself inside a Josephson junction. Most importantly, at what F-layer thickness, $d_F$, does the junction transition from the 0 state to the $\pi$ state? Secondly, how does $I_c$ decay as $d_F$ increases? Thirdly, how much does the field generated by the magnetization affect the junction properties? And lastly, does the material exhibit single-domain switching behavior when it is patterned into a micron-scale elliptical nanomagnet? All of those questions can be answered by fabricating and measuring micron-scale SFS junctions containing only a single magnetic layer~\cite{Niedzielski2015}; the properties of the spin-valve junctions should then be predictable based on the results from the single-layer junctions. In this work, we report on junctions containing either Ni$_{65}$Fe$_{15}$Co$_{20}$ or Ni$_{81}$Fe$_{19}$. Our motivation for studying NiFeCo was explained above. While NiFe has already been studied by other groups \cite{Robinson2007,Qader2014}, there is no guarantee that we can safely rely on those previous results. The 0-$\pi$ transition thickness of NiFe was reported to be 1.2 nm by Robinson \textit{et al.}\cite{Robinson2007} and 2.3 nm by Qader \textit{et al}\cite{Qader2014}. Since we continue to use NiFe in our spin-valve junctions, it is important for us to characterize the NiFe grown in our lab.
	
	\section{\label{sec:level2}Sample Fabrication and Characterization}
	Our SFS Josephson junctions are fabricated using ultrahigh-vacuum sputtering deposition on 0.5$\times$0.5 in$^2$ silicon chips. The geometry of the bottom leads was defined via optical photolithography and the positive photoresist S1813.
	
	Before the sputtering deposition, the chamber was baked for eight hours and reduced to a base pressure of 2$\times$10$^{-8}$ Torr with a cryopump. The chamber was then cooled by circulating liquid nitrogen though a Meissner trap to reduce the partial pressure of water in the chamber to $<$ 3 $\times$ 10$^{-9}$ Torr as confirmed by an \textit{in-situ} residual gas analyzer. The films were deposited via dc sputtering in an Argon plasma with pressure 1.3 $\times$ 10$^{-3}$ Torr. During deposition the sample temperature was held between $-30\,^{\circ}\mathrm{C}$ and $-20\,^{\circ}\mathrm{C}$ measured with a thermocouple affixed to the back of one of the substrates.
	
	In a single run, a rotating sample plate and shutter system passes up to 16 chips over a sequence of triode sputtering guns containing 2.25-inch diameter targets of Nb, Al, NiFe, NiFeCo, and magnetron guns containing 1-inch diameter targets of Au and Cu. The thicknesses of the various deposited materials were controlled by measuring the deposition rates (accurate to $\pm 0.1 \mathrm{\AA}$/s) using a crystal film thickness monitor and a computer controlled stepper motor that operates the position of the shutter and sample plate.
	
	\begin{figure}
		\begin{center}
			\includegraphics[width=2.8 in]{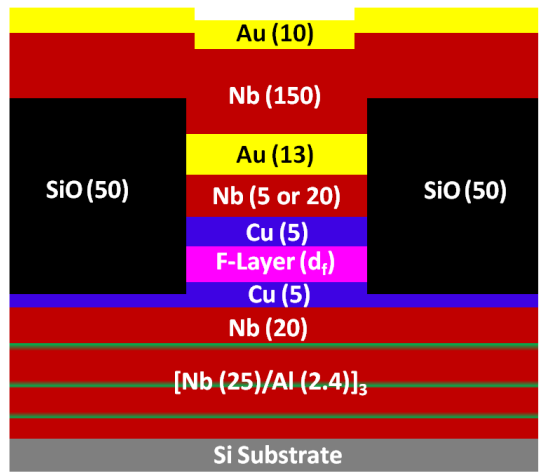}
		\end{center}
		\caption{\label{fig:SampleStructure} A schematic representation of the vertical cross-sectional structure of our SFS Josephson junctions. The F layer is either Ni$_{65}$Fe$_{15}$Co$_{20}$ or Ni$_{81}$Fe$_{19}$ with thickness $d_F$ ranging from 0.8 to 3.8 nm. All other units are given in nanometers.}
	\end{figure}

	\begin{figure}
		\begin{center}
			\includegraphics[width=3.4 in]{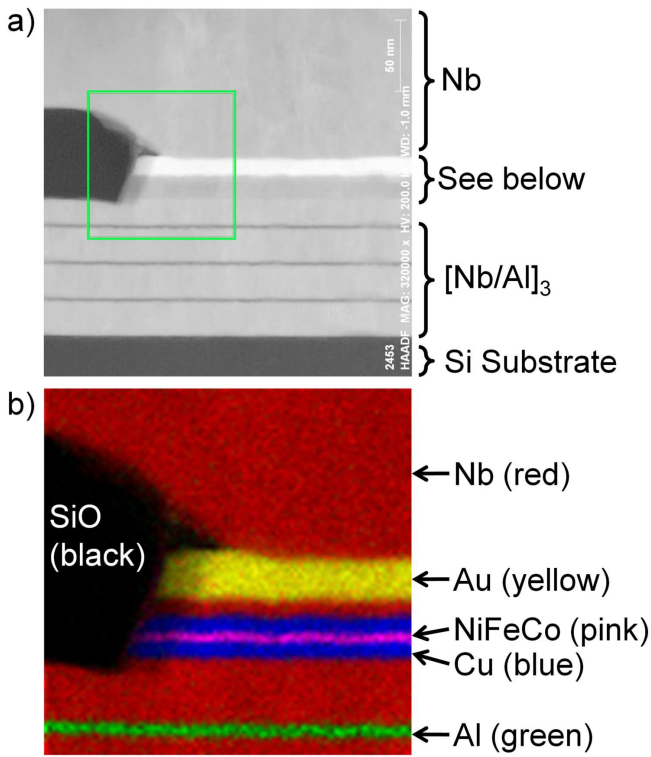}
		\end{center}
		\caption{\label{fig:TEMImage} Vertical cross-sections through the center of our SFS Josephson junctions are prepared with focused ion beam (FIB) milling and imaged with Scanning Transmission Electron Microscopy (STEM) to validate the fabrication process. In (a) the STEM image shows where our [Nb/Al] bottom electrode meets the left edge of the patterned junction area and SiO barrier. Energy dispersive x-ray spectroscopy (EDX) was used to map out the elemental composition of the individual layers within. In (b) we show the EDX analysis for the region outlined by the green square, which clearly shows a continuous NiFeCo layer ($d_F$=1.6 nm).}
	\end{figure}
	
	The bottom wiring layer, which was deposited without breaking vacuum, consists of the sequence [Nb(25)/Al(2.4)]$_3$/Nb(20)/Cu(5)/F($d_F$)/Cu(5)/Nb(5 or 20)/Au(10), with thicknesses given in nanometers. A schematic of the sample structure is shown in Fig.~\ref{fig:SampleStructure}. In order to verify the fabrication process, cross-sectional areas of the junctions were investigated by high-resolution scanning transmission electron microscopy (STEM) and energy dispersive x-ray spectroscopy (EDX). The cross-sections were prepared using a FEI Helios focused ion beam (FIB) with a Ga ion source, and transferred to a Ti grid for imaging in a FEI Titan G2 80-200 aberration-corrected STEM operated at 200kV and equipped with four silicon drift X-ray detectors. The high-angle annular dark field STEM image in Fig.~\ref{fig:TEMImage}(a) shows the left hand side of one of the NiFeCo junction stacks.

	To achieve quality magnetic switching it is crucial to grow the ferromagnets on a smooth underlayer. The low surface roughness of the [Nb/Al] multilayer~\cite{Wang2012, Thomas1998, Kohlstedt1996} provides a smooth template for subsequent growth of the Cu spacer and ferromagnetic layer, where the Al is thin enough to be superconducting through the proximity effect with Nb. We independently measured the roughness of [Nb/Al] multilayer to be $\approx 2.3$ $\mathrm{\AA}$ using atomic force microscopy (AFM), smoother than a single Nb(100) film ($>5$ $\mathrm{\AA}$). Significant diffusion of Al along Nb grain boundaries was not observed. The superconducting transition temperature of [Nb(25)/Al(2.4)]$_3$/Nb(20) films, as measured with a SQUID magnetometer, is 8.0 K, well above the temperature at which we measure our SFS junctions (4.2 K).
	
	Due to the crystal lattice mismatch between the fcc ferromagnetic materials and the bcc Nb we add a 5-nm Cu spacer on either side of the F layer. STEM diffraction patterns show that the Cu layer grows with a [111] orientation on Nb [011]. Grains with favorable orientation relative to the beam direction show lattice fringes extending through the entire Cu/ferromagnetic layer/Cu thickness. In comparison to films grown on only Nb the spacer improves the magnetic properties of the F layers: decreasing the coercive field, and increasing the sharpness of hysteresis loops. Also, smooth normal metal spacer layers will be used in cryogenic memory to magnetically decouple the multiple F layers.
	
	EDX phase maps were created by performing a multivariate statistical analysis of the spectra from each individual pixel, and color-coding pixels containing the same spectrum~\cite{Kotula2006}. The phase map shown in Fig. 2b corresponds to the area within the green square in Fig. 2a. The 1.6 nm NiFeCo layer is clearly uniform and continuous, consistent with the magnetic behavior discussed below.
	
	To set the direction of the magnetocrystalline anisotropy of the ferromagnetic alloys, the NiFeCo samples were sputtered in a magnetic-field of $\approx$ 80 mT (whose direction points along what will become the major axis of our elliptical junctions) generated by small NdFeB magnets affixed directly on the back of the substrates. The NiFe samples, made in separate sputtering runs, were sputtered in a magnetic-field of $\approx$ 50 mT. Finally the samples were capped with a thin layer of Nb and Au to prevent oxidation.
	
	The junctions were patterned by electron-beam lithography followed by ion milling in Argon. We use the negative e-beam resist ma-N2401 as the ion mill mask. The junctions are elliptical in shape with an aspect ratio of 2.5 and area of either 0.1, 0.25 or 0.5 $\mu$m$^2$, all sufficiently small to ensure that the magnetic layer is single domain. Elliptically-shaped junctions have the advantages that the modulation of the critical current through the junction versus the applied magnetic field, known as a Fraunhofer pattern, follows an analytical formula while the (small) demagnetizing field is nearly uniform when the magnetization is uniform.\footnote{Strictly speaking, the field is uniform only inside a uniformly magnetized ellipsoid. Because the elliptical nanomagnets in our junctions are very thin, there is very little difference between an ellipse and an ellipsoid.}
	
	Outside the mask region, we ion milled though the capping layer, the F layer, and nominally half-way into the underlying Cu spacer layer. Figure~\ref{fig:TEMImage} confirms our patterning of the F layer, though it is clear this sample was slightly over-milled; the step edge extends through the second Cu layer and just into the underlying Nb. After ion milling, a 50 nm thick SiO layer was deposited by thermal evaporation to electrically isolate the junction and the bottom wiring layer from the top wiring layer. During thermal evaporation the sample was rotated to improve the uniformity of the oxide and reduce pinhole formation.
	
	To prevent the e-beam resist from over-heating during the ion milling and SiO deposition the back of the substrate is placed against a Cu heatsink with a small drop of vacuum grease or silver paste to improve thermal contact. A capping layer containing 20 nm Nb was used in some of the NiFe-based samples, but was later reduced to 5 nm in the remaining NiFe and NiFeCo-based samples for two main reasons: i) During ion milling a veil of Nb can be backsputtered onto the edge of the e-beam resist, preventing the e-beam resist from lifting-off properly, or at all. Reducing the Nb thickness reduces the extent of the veil. ii) Since Nb has the slowest etching rate of all our materials, reducing the Nb thickness drastically reduces the total time required to ion mill. Reducing the Nb thickness in the capping later improved our lift-off success rate, likely due to less damage and distortion of the resist under the heat of the ion mill and during SiO deposition.
	
	Finally, the top Nb wiring layer was patterned using the same photolithography and lift-off process as the bottom leads. Residual photoresist on the surface is cleaned with oxygen plasma etching followed by \textit{in-situ} ion milling in which 2 nm of the top Au surface is removed prior to sputtering. We deposited top leads of Nb(150 nm)/Au(10 nm), ending with the Au to prevent oxidation.
	
	\section{\label{sec:level3}Measurement and Analysis}
	\begin{figure}
		\begin{center}
			\includegraphics[width=3.4 in]{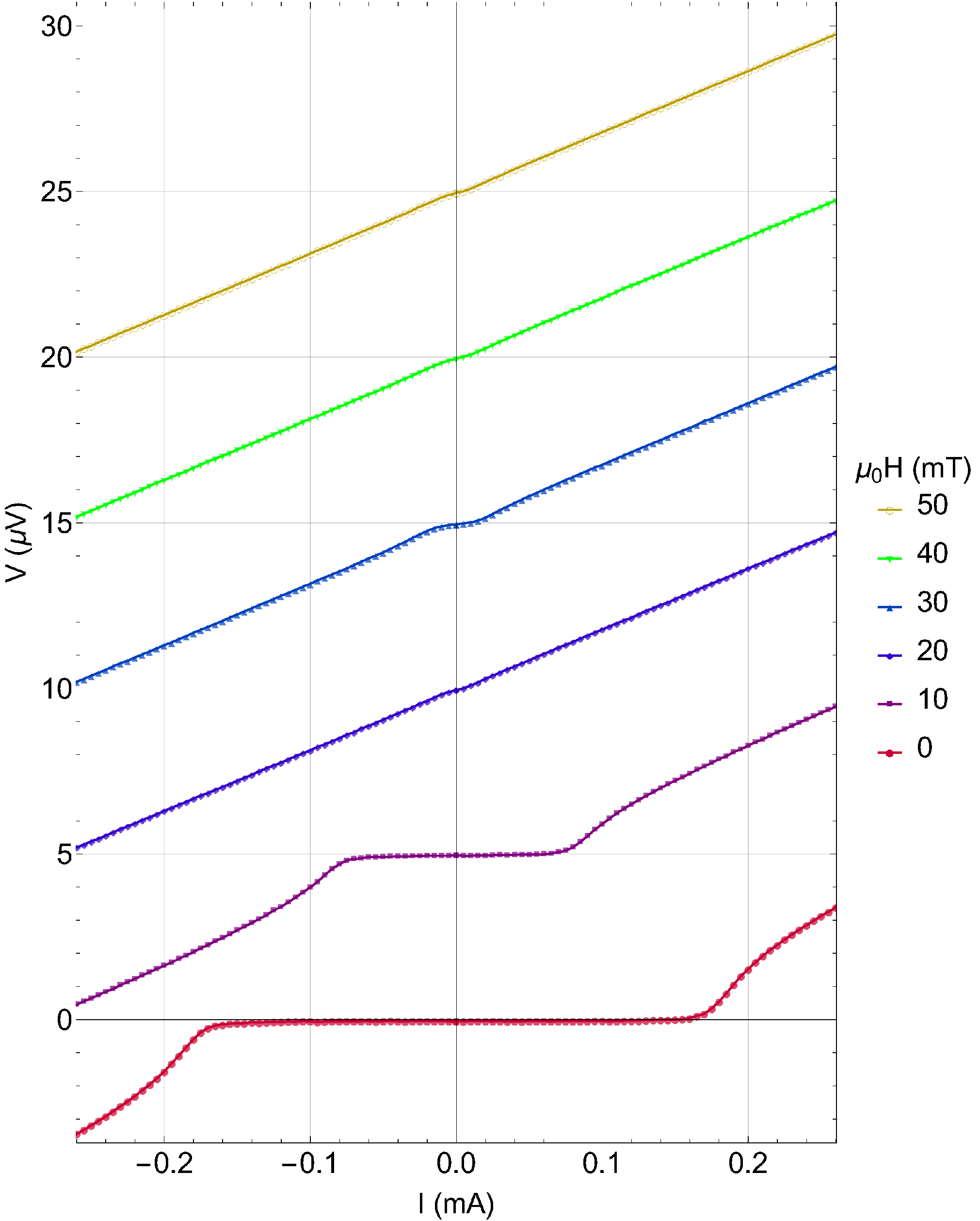}
		\end{center}
		\caption{\label{fig:IVcurve} The voltage across an overdamped SFS Josephson junction containing 2 nm layer of NiFeCo, versus the applied current. The data are measured via standard four-terminal measurement in an external magnetic field of 0 - 50 mT as indicated. The critical current I$_c$, extracted from I-V curves above, is used to produce the Fraunhofer pattern shown in Fig. \ref{fig:NiFeCo_Fraunhofers} (b). For clarity, each successive curve is shifted along the voltage axis in steps of 5 $\mu$V.}
	\end{figure}

	\begin{figure}
		\begin{center}
			\includegraphics[width=3.2 in]{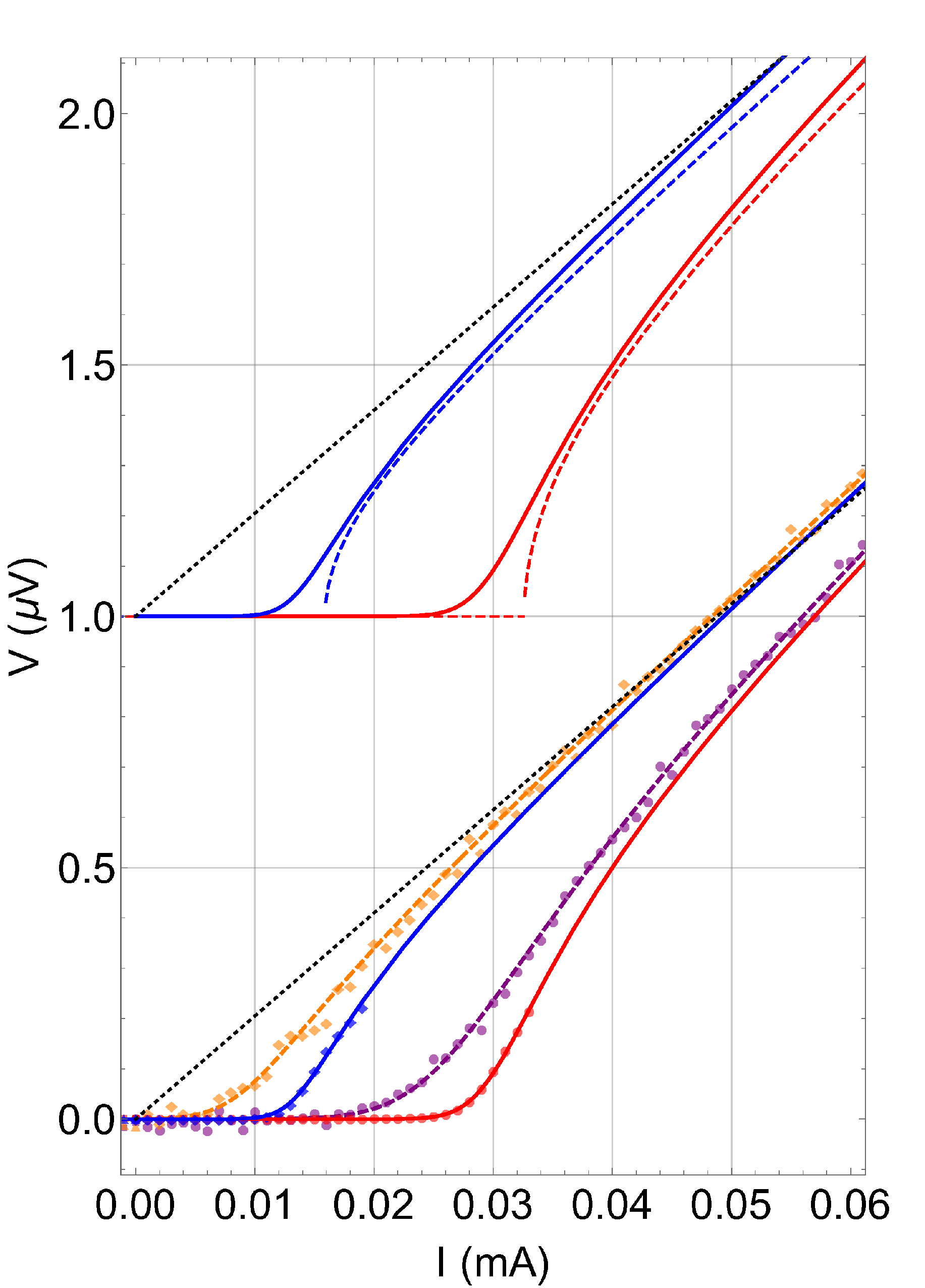}
		\end{center}
		\caption{\label{fig:IZfit} The voltage across an SFS Josephson junction versus the applied current at applied magnetic fields of 18 mT (circles) and 36 mT (diamonds), using two different measurement schemes. The orange and purple colored data points are measured with a commercial nanovoltmeter while the blue and red colored data points are measured with an rf SQUID in a self-balancing potentiometer comparator circuit. The data are fit to the Ivanchenko-Zil'berman function, Eqn. \ref{eq:IvachenkoZilbermanEqn}, which accounts for noise-induced rounding and allows us to extract the noise temperature of our measurement systems. The upper set of curves (shifted along the voltage axis by 1 $\mu$V for clarity) compares the two different fitting methods: to the square root function (Eqn. \ref{eqn:SquareRootFunction}, dashed lines) and to the Ivanchenko-Zil'berman function (Eqn. \ref{eq:IvachenkoZilbermanEqn}, solid lines). The dotted black line represents Ohms' law for the measured normal state resistance. This junction contains a 1 nm layer of NiFeCo.}
	\end{figure}
	
	Each device was connected to the wire leads of a dip-stick probe with pressed indium solder. The samples were immersed in a liquid-He dewar outfitted with a Cryoperm magnetic shield and placed inside a shielded room to reduce noise from external sources of electromagnetic radiation. The dipping probe is equipped with a superconducting solenoid used to apply uniform magnetic fields over a range of \mbox{-60} to 60 mT along the long-axis of the elliptical junctions in the plane of the sample. The current-voltage characteristics of the junctions were measured at 4.2 K in a four-terminal configuration with one or both of the following methods: 1) a Yokogawa current source provides a bias current to the Josephson junction while voltage measurements were made with a Keysight nanovoltmeter or 2) a commercial Quantum Design rf SQUID in a self-balancing potentiometer comparator circuit measures the voltage across the junction, while the measurement current is provided by a battery-powered ultra-low noise programmable current source~\cite{Edmunds1980}. Data taken independently from the two setups agree with one another, however the rf SQUID comparator scheme has lower RMS voltage noise of only 6 pV compared to 11 nV for the commercial nanovoltmeter, with both systems measuring over 10 power line cycles. Typical I-V curves, shown in Figs.~\ref{fig:IVcurve} and \ref{fig:IZfit}, have the expected behavior of overdamped Josephson junctions~\cite{Barone1982}. Figure~\ref{fig:IVcurve} shows how the I-V curve shape changes while being measured in applied magnetic fields ranging from 0-50 mT. The entire data collection process is automated using the LabView software package.
	
	The sample resistance in the normal state $R_N$ was determined by the slope of the linear region of the I-V curve when $|I| \gg I_c$. While the sensitivity of the rf SQUID measurement system allows us to measure junctions with small I$_c$, it operates only over a restricted voltage range. Thus, depending on the resistance of the sample, one may be limited in the extent to which the linear tail of the I-V curve can be measured. In these cases, independent measurements using both measurement schemes are necessary to accurately determine both I$_c$ and R$_N$. Measurements of the area-resistance product in the normal state yield consistent values of $A R_N$ = 5-10 f$\Omega$-$m^2$ for both the NiFeCo and NiFe based junctions of different areas -- an indicator of the reproducible high quality interfaces. This total specific resistance is close to twice the Nb/F interface resistance, determined in separate current-perpendicular-to-plane giant-magnetoresistance studies~\cite{Vila2000}.
	
	The critical current $I_{c}$ was extracted by fitting the I-V curves to a square root function of the form,
	\begin{equation}
	\label{eqn:SquareRootFunction}
	V= R_N \sqrt{I^2 - I_c^2}, \hspace*{0.1in} I \ge I_c.
	\end{equation}
	We occasionally observe that $I_c$ is slightly different in the positive and negative current directions. That does not violate any symmetry given the presence of the ferromagnetic material in the junctions, but we find it puzzling given the rather small value of mutual inductance between the electrical leads and the junction proper. In those cases we define $I_c$ to be the average value of the critical currents in the two directions.
	
	When the critical current of the junctions is small, there is a noticeable amount of rounding of the I-V curves as $I$ approaches $I_c$.  Ivanchecko and Zil'berman (IZ) and Ambegaokar and Halperin developed a theory to fit such data when the rounding is due to thermal fluctuations~\cite{IvanchenkoZilberman1969, AmbegaokarHalperin1969}.  (In the “tilted-washboard” potential of the RCSJ model, thermal fluctuations cause the particle to escape out of the potential wells when $I < I_c$.)  When the rounding is caused by fluctuations in the electromagnetic environment coupled to the junction (usually from the measurement apparatus), the temperature in the IZ model becomes an effective temperature, which can be much larger than the actual sample temperature. In the IZ model the I-V curve has the analytical solution~\cite{IvanchenkoZilberman1969}
	\begin{equation}
	\label{eq:IvachenkoZilbermanEqn}
	V(I_c, R_N, T_{\mathrm{eff}}) = I_c R_N \left( \frac{I}{I_c} - \mathcal{I}_{-} + \mathcal{I}_{+} \right), \hspace*{0.1in} I \ge 0,
	\end{equation}
	where $\mathcal{I}_{\pm} = \Big(\!\frac{\mathcal{I}_{(1 \pm i) \gamma }(\gamma_c)}{2i\mathcal{I}_{\pm i\gamma}(\gamma_c)} \!\Big)$, $\gamma =I \hbar /(2 e k_B T_{\mathrm{eff}})$ and $\gamma_c= I_c \hbar /(2 e k_B T_{\mathrm{eff}})$. $\mathcal{I}_{\nu}$(z) are modified Bessel functions of the first kind with $\nu$ a non-integer complex number, where $e$ is the electron charge, $k_B$ is the Boltzmann constant, and $T_{\mathrm{eff}}$ is the effective noise temperature.
	
	Figure~\ref{fig:IZfit} shows fits of the IZ function to data from a sample at magnetic fields where $I_c$ is rather small, hence the rounding is apparent.  Data are presented both for the nanovoltmeter-based measurement system and for the rf-SQUID-based system.  The noise temperature, $T_{\mathrm{eff}}$, is $\approx$ 95 K for the former, versus $\approx$ 37 K for the latter, as shown in Table~\ref{table:IZfitparams}. The table shows that the values of $I_c$ extracted from the fits are comparable despite the difference in $T_{\mathrm{eff}}$. However, due to it's much lower RMS voltage noise, the SQUID-based measurement system was used for samples whose maximum value of $I_c$ is less than about 10 $\mu A$.
	
	\begin{table}
		\caption{\label{table:IZfitparams} Parameters obtained from fits of the Ivanchenko-Zil'berman function to the data shown in Fig. \ref{fig:IZfit}. The normal-state resistance R$_N$ was measured to be 20.5 m$\Omega$, and was not used as a free fitting parameter. }
		\begin{tabular}{ccc@{\hskip 0.25in}c}
			{Method} & {Field (mT)} & {$I_c$ (mA)} & {$T_{\mathrm{eff}}$ (K)} \\ \hline\hline \noalign{\smallskip}
			nanovoltmeter & 18 & 0.0310 $\pm$ 1E-4 & 98.6 $\pm$ 4.8  \\
			& 36  & 0.0126 $\pm$ 3E-4 & 85.3 $\pm$ 12.9  \\ \hline \noalign{\smallskip}
			rf SQUID & 18  & 0.0326 $\pm$ 5E-5  & 37.4 $\pm$ 0.7  \\
			& 36  & 0.0159 $\pm$ 1E-4 & 36.5 $\pm$ 2.8 \\ \hline\hline\noalign{\smallskip}
			& & & \\
		\end{tabular}
		\setlength{\tabcolsep}{12pt}
	\end{table}

	\begin{figure}
		\begin{center}
			\includegraphics[width=3.2 in]{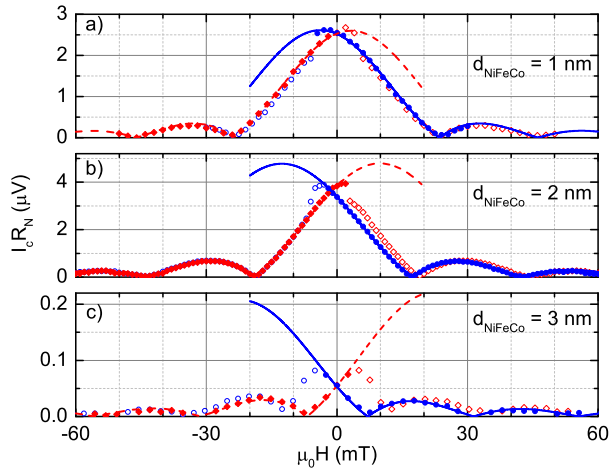}
		\end{center}
		\caption{\label{fig:NiFeCo_Fraunhofers} NiFeCo Fraunhofer patterns: Critical current times the normal state resistance, $I_c R_N$, is plotted versus the applied field $H$, for three samples with $d_{\mathrm{NiFeCo}}$= (a) 1 nm, (b) 2 nm, (c) 3 nm. The data before $H_{\mathrm{switch}}$, the field at which the NiFeCo magnetization reverses direction (solid markers), and the corresponding fits (lines) to Eqn.~\ref{eqn:FraunhoferAiryFit} show excellent agreement for both the positive (red, dashed) and negative (blue) field sweep directions. The hollow circles are the corresponding data points after $H_{\mathrm{switch}}$. The Fraunhofer patterns display magnetic hysteresis and are increasingly shifted with larger $d_{\mathrm{NiFeCo}}$.}
	\end{figure}
	
\begin{figure}
		\begin{center}
			\includegraphics[width=3.2 in]{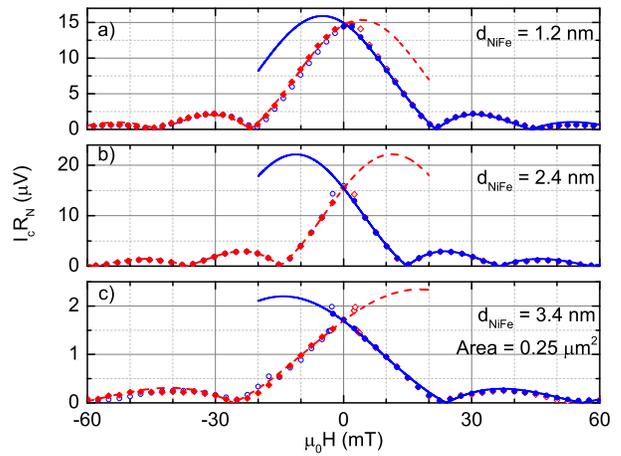}			
		\end{center}
		\caption{\label{fig:NiFe_Fraunhofers} NiFe Fraunhofer patterns: $I_c R_N$ vs applied field $H$, varying $d_{\mathrm{NiFe}}$ from (a) 1.2 nm, (b) 2.4 nm, and (c) 3.4 nm. The nominal junction area in (a) and (b) is 0.5 $\mu$ m$^2$, and in (c) is 0.25 $\mu$ m$^2$, as evidenced by (c)'s comparatively wider Fraunhofer pattern.}
	\end{figure}

	Measuring $I_c$ as a function of the applied magnetic field, we map out so-called ``Fraunhofer'' diffraction patterns, shown in Figs.~\ref{fig:NiFeCo_Fraunhofers} and \ref{fig:NiFe_Fraunhofers} for F = NiFeCo and NiFe respectively. To compare junctions with different cross-sectional areas we normalized our data by multiplying $I_c$ by $R_N$. For elliptical junctions the functional form is an Airy function~\cite{Barone1982},
	\begin{equation}
	\label{eqn:FraunhoferAiryFit}
	I_{c}=I_{c0} \left| 2 J_{1} \left( \pi \Phi / \Phi_{0} \right) / \left( \pi \Phi / \Phi_0 \right) \right|,
	\end{equation}
	\added[id=JAG]{where $J_1$ is an unmodified Bessel function of the first kind (whose order is a real integer, unlike the modifed Bessel functions of Eq.~\ref{eq:IvachenkoZilbermanEqn})} and $I_{c0}$ is the maximum critical current and $\Phi_0 = h/2e$ is the flux quantum. The magnetic flux through the junction is \cite{Khaire2009}\footnote{We correct a missing factor of $\mu_0$ to the corresponding equation in Ref.~\onlinecite{Khaire2009}},
	\begin{equation}
	\label{eqn:magneticflux}
	\Phi=\mu_0 H w (2 \lambda_L+2d_N+d_F) + \mu_0 M w d_F,
	\end{equation}
	where $H$, $w$, $\lambda_L$, $d_N$ and $d_F$ are the applied field, the width of the junction (minor axis), the London penetration depth of the Nb electrodes, and the thicknesses of the normal metal and F layers. The last term in Eqn.~\ref{eqn:magneticflux} arises from the magnetization $M$ of the single-domain ferromagnet, and is valid only if $M$ is uniform and points along the same direction as the applied field $H$. Eqn.~\ref{eqn:magneticflux} neglects the small contributions to $\Phi$ from the uniform demagnetizing field and any magnetic field from the nanomagnet that returns inside the junction. From Eqn.~\ref{eqn:magneticflux} it is clear that the Fraunhofer pattern will be shifted along the field axis by an amount $H_{\mathrm{shift}}= -M d_F/(2 \lambda_L+ d_F+ 2 d_{\mathrm{Cu}})$.
	
	The data shown in Figs.~\ref{fig:NiFeCo_Fraunhofers} and \ref{fig:NiFe_Fraunhofers} were acquired as follows: \added[id=JAG]{we} first applied a field of -60 mT to fully magnetize the nanomagnet, then the field was slowly ramped to +60 mT in steps of typically 1.5 mT. At a critical value $H_{\mathrm{switch}} >$ 0, the ferromagnet undergoes a rapid reversal of it's magnetization direction. The data then transition to another Fraunhofer pattern shifted in the opposite direction. The applied magnetic field was then swept in the reverse orientation to observe the magnetic hysteresis. Similar magnetic hysteresis loops in Josephson junctions have been previously studied \cite{Baek2014,Niedzielski2015}, and are well understood.
	
	We fit the data starting from the initialization field to $H_{\mathrm{switch}}$ with Eqn.~\ref{eqn:FraunhoferAiryFit}, where $I_{c0}$ and $H_{\mathrm{shift}}$ are fitting parameters. Allowing the sample width, $w$, to be a free fitting parameter gives rise to large uncertainty in $H_{\mathrm{shift}}$ for data sets with large values of $H_{\mathrm{shift}}$. Hence we fixed $w$ to its nominal value in all the fits presented here. We keep $\lambda_L$ fixed at 85 nm, as determined by data obtained in our group over many years \cite{Khaire2009}. In Figs.~\ref{fig:NiFeCo_Fraunhofers} and \ref{fig:NiFe_Fraunhofers} the corresponding fits (lines) show excellent agreement with the fitted data (solid markers), for the positive (red) and negative (blue) sweep directions. The hollow data points after $H_{\mathrm{switch}}$, whose fits are not plotted for clarity, match well with the Fraunhofer pattern from where the field is swept in the opposite direction. The nodes in the Fraunhofer patterns closely approach $I_c=0$, which illustrates that there are no shorts in the SiO. This and the lack of distortion in the pattern at large field values indicates that there is little if any trapped flux in the junction electrodes.
	
	The excellent quality of the Fraunhofer patterns starting at high field and extending past zero field to $H_{\mathrm{switch}}$ indicates that the remanent magnetization in the junctions is uniform, suggesting that the ferromagnetic layers are probably single-domain.  In many cases, the magnetic switching is abrupt, which also supports that interpretation. The excellent quality of the fits allows us to extract the peak value of $I_c$ even in cases where the peak is inaccessible in the data because it lies beyond the field where the magnetic layer switches. The uncertainties in peak height and position in such cases, are, of course, larger than when the peak is directly visible in the data.
	
	Some Fraunhofer patterns, such as those shown in Figs.~\ref{fig:NiFeCo_Fraunhofers} (b)-(c), show that the reversal of the magnetization occurs over a range of field values, implying that our junctions do not strictly follow the abrupt switching behavior predicted by the Stoner-Wolfarth model. For those samples, magnetization reversal takes place through a non-uniform intermediate state, for example an ``S-shaped" state or a multi-domain state. This could be exacerbated by a number of factors including nonuniform magnetocrystalline anisotropy, surface or edge roughness, or magnetostriction.
	
	\begin{figure}
		\begin{center}
			\includegraphics[width=3.0 in]{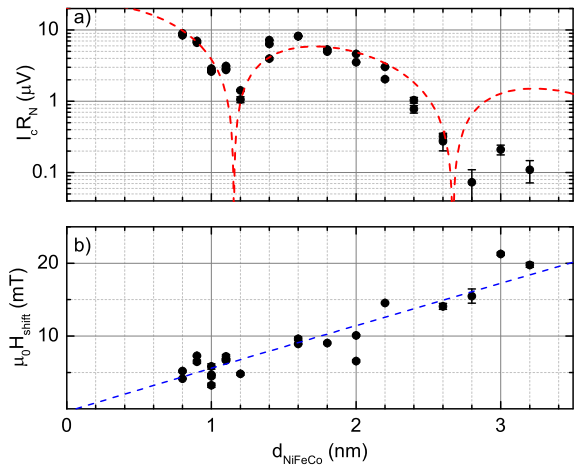}
		\end{center}
		\caption{\label{fig:IcRn_HShift_NiFeCo} a) The maximal $I_c$ times $R_N$ is plotted versus $d_{\mathrm{NiFeCo}}$ for many samples, with the error bars determined by the goodness of fit parameters of the individual Fraunhofer patterns. The data are fit to Eqn.~\ref{eq:IcRnZeroPiTransition}, and the best fit parameters are shown in Table \ref{table:0Pifitparams}. The first minima indicates that at a critical NiFeCo thicknesses of 1.15 $\pm$ 0.02 nm there is a transition between the 0 and $\pi$-phase states. b) The Fraunhofer pattern field shift $H_{\mathrm{shift}}$ is shown to increase with $d_{\mathrm{NiFeCo}}$. The fit to Eqn.~\ref{eqn:Hshift} yields a magnetization of 855 $\pm$ 81 kA/m and an x-intercept which corresponds to no discernible magnetic dead layer ($d_\mathrm{dead}$=0.06 $\pm$ 0.17 nm).}
	\end{figure}
	
\begin{figure}
		\begin{center}
			\includegraphics[width=3.0 in]{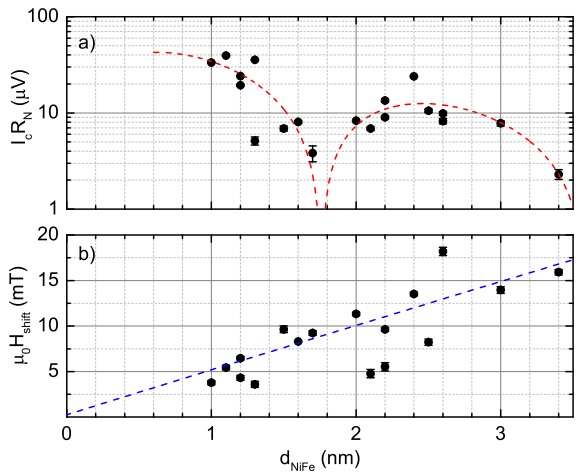}
		\end{center}
		\caption{\label{fig:IcRn_Hshift_NiFe} a) The maximal $I_c$ times $R_N$ is plotted versus $d_{\mathrm{NiFe}}$ for many samples, with the error bars determined by the goodness of fit parameters of the individual Fraunhofer patterns. The data are fit to Eqn.~\ref{eq:IcRnZeroPiTransition}, and the best fit parameters are shown in Table \ref{table:0Pifitparams}. The first minima indicates that at a critical NiFe thicknesses of 1.76 $\pm$ 0.05 nm the junctions transition between the 0 and $\pi$-phase states. b) Despite some scatter the Fraunhofer pattern field shift $H_{\mathrm{shift}}$ increases with $d_{\mathrm{NiFe}}$. The fit to Eqn.~\ref{eqn:Hshift} yields a magnetization of 711 $\pm$ 144 kA/m and no discernible magnetic dead layer ($d_\mathrm{dead}= -$0.05 $\pm$ 0.55 nm).}
	\end{figure}

	The thickness at which the junctions transition from a 0 to $\pi$-phase state is determined by plotting $I_c R_N$ for many samples spanning a range of ferromagnet thicknesses $d_F$, and looking for deep local minima where $I_c$ theoretically passes through zero. This is shown in Figs.~\ref{fig:IcRn_HShift_NiFeCo}(a) and \ref{fig:IcRn_Hshift_NiFe}(a) where $I_c$ denotes the maximum critical current obtained from the Fraunhofer pattern fits. The transition from a 0 to $\pi$-phase state occurs at thicknesses of about $d_{\mathrm{NiFeCo}}$=1.2 nm and $d_{\mathrm{NiFe}}$=1.8 nm. The latter value is between the values found by Robinson \textit{et al.} \cite{Robinson2006,Robinson2007} and by Qadar \textit{et al.} \cite{Qader2014}.
	
	Theoretical predictions for the behavior of $I_c R_N$ versus thickness of the ferromagnetic layer describe an oscillating function with either an algebraic decay for ballistic transport~\cite{Buzdin1982} or an exponential decay for diffusive transport~\cite{Buzdin1991}. The crossover from the ballistic to the diffusive limit has also been addressed in several theoretical works\cite{Bergeret2001, Pugach2011}. In the diffusive limit the behavior is governed by the Usadel equations~\cite{Buzdin_SFReview2005} in cases where the majority and minority spin bands have nearly identical properties. For strong ferromagnetic materials, the Usadel equations are not adequate~\cite{Bergeret2001}. In principle, one should take into account the Fermi-surface mismatch at each interface, as well as the different densities of states, mean free paths, and diffusion constants for the majority and minority spin bands. Microscopic calculations based on the Bogoliubov-deGennes equations and taking into account the finite interface transparency, have been performed for ballistic systems~\cite{Radovic2003, Barsic2007}, and could, in principle, be extended to systems with disorder. But there have been no theoretical calculations of the supercurrent that take into account the complex band structure of transition-metal ferromagnetic materials such as those discussed in this paper. Nevertheless, several previous experimental works have used existing theoretical formulas to fit data from strong ferromagnetic materials. For instance, the ballistic form was used by Robinson \textit{et al.} to fit data from junctions containing elemental ferromagnets, Ni, Co, and Fe, but data from junctions containing NiFe appeared to show a crossover from ballistic to diffusive behavior at a thickness of about 2 nm~\cite{Robinson2006, Robinson2007}. We also attempted to fit our data to the ballistic limit used in Ref.~\onlinecite{Robinson2007}, but did not find good agreement.
	
	The data shown in Figs.~\ref{fig:IcRn_HShift_NiFeCo}(a) and \ref{fig:IcRn_Hshift_NiFe}(a) roughly follow an exponential decay, which is not surprising given the short mean free paths of minority carriers in NiFe and NiFeCo~\cite{Bass2016}. It should be emphasized however that with our thin F layers, the concept of a mean free path may not be a valid notion when considering that the dominant scattering occurs at the F-layer/Cu interfaces. We estimate that ratio of interfacial minority-majority scattering for our F-layer materials to be $\approx$ 6 \cite{Bass2016}. Regardless, we fit the data for both NiFeCo and NiFe to the diffusive limit case with the function,
		\begin{equation}
		\label{eq:IcRnZeroPiTransition}
		I_c R_N = V_0 e^{-d_F/ \xi_{F1}} \Big| \sin \left( \frac{d_F - d_{0-\pi}}{\xi_{F2}} \right) \Big|,
		\end{equation}
		where $\xi_{F1}$ and $\xi_{F2}$ are length scales that control the decay and oscillation period of $I_c$ with $d_F$, and $d_{0-\pi}$ is the 0-$\pi$ transition thickness. The simplest model of a diffusive S/F/S Josephson junction based on the Usadel equation \cite{Buzdin_SFReview2005} predicts that $\xi_{F1} = \xi_{F2} = \sqrt{\hbar D_F/E_{ex}}$ where $D_F$ and $E_{ex}$ are the diffusion constant and exchange energy of F, respectively, and has an overall phase offset $\phi \equiv (d_{0-\pi} / \xi_{F2}) - \pi/2 = \pi/4$. In the presence of spin-orbit or spin-flip scattering~\cite{Faure2006}, or when the F-layer contains domain walls~\cite{Bakurskiy2015}, one expects to find $\xi_{F1} < \xi_{F2}$. \added[id=NOB]{For strong ferromagnetic materials with large exchange energy, however, one often finds that $\xi_{F1} > \xi_{F2}$ \cite{Robinson2007}, a result that can be explained from a semi-ballistic calculation starting from the Eilenberger equation\cite{Bergeret2001, Pugach2011}.}  In addition, the phase offset, $\phi$, has been shown by Heim \textit{et al.}~\cite{Heim2015} to vary sensitively with the type and thickness of normal-metal spacer layers or insulating barriers included in the junction.
	
	\begin{table}
		\caption{\label{table:0Pifitparams} Best-fit parameters from Eq.~\ref{eq:IcRnZeroPiTransition} corresponding to the data in Figs.~\ref{fig:IcRn_HShift_NiFeCo}(a) and ~\ref{fig:IcRn_Hshift_NiFe}(a). }
		\begin{tabular}{c@{\hskip 0.14in}c@{\hskip 0.14in}c@{\hskip 0.14in}c@{\hskip 0.14in}c}
			{F-layer} & {$d_{0-\pi}$ (nm)} & {$V_0$ ($\mu$V)} & {$\xi_{F1}$ (nm)} & {$\xi_{F2}$ (nm)} \\ \hline\hline \noalign{\smallskip}
			NiFeCo & 1.15 $\pm$ 0.02  & 30 $\pm$ 6  & 1.11 $\pm$ 0.16 & 0.48 $\pm$ 0.03 \\ \hline \noalign{\smallskip}
			NiFe& 1.76 $\pm$ 0.05  & 69 $\pm$ 19  & 1.50 $\pm$ 0.38 & 0.58 $\pm$ 0.10 \\ \hline\hline
			& & & &
		\end{tabular}
		\setlength{\tabcolsep}{12pt}
	\end{table}
	
	Equation~\ref{eq:IcRnZeroPiTransition} fits the data reasonably well for both our NiFeCo and NiFe based junctions.  Table \ref{table:0Pifitparams} lists the best-fit parameters for the data in Figs. \ref{fig:IcRn_HShift_NiFeCo}(a) and \ref{fig:IcRn_Hshift_NiFe}(a), and shows that, in spite of the significant scatter in the data, the thickness corresponding to the first 0-$\pi$ transition can be extracted with reasonable precision. Despite this, Eqn.~\ref{eq:IcRnZeroPiTransition} does not fit so well for the the thickest subset of the NiFeCo samples, where one would have expected a better fit in this more-diffusive regime. Both types of junctions appear to have $\xi_{F1} > \xi_{F2}$, and in the case of NiFe, the values are similar to those found by Robinson \textit{et al.} ($\xi_{F1}= 1.4$ nm, $\xi_{F2}= 0.46$). Combining our results with those of Robinson \textit{et al.}, one might conclude that $\xi_{F1} > \xi_{F2}$ is a generic condition for Josephson junctions containing strong transition-metal ferromagnetic materials. That is not true, however, as the thickness dependence of $I_c R_N$ in junctions containing NiFeMo was fit very well by Eqn.~\ref{eq:IcRnZeroPiTransition} but with $\xi_{F1} < \xi_{F2}$, presumably due either to the very short mean free path or very short spin diffusion length in that material \cite{Niedzielski2015}.
	
	In Figs.~\ref{fig:IcRn_HShift_NiFeCo}(b) and ~\ref{fig:IcRn_Hshift_NiFe}(b), for each junction we plot the average of the $H_{\mathrm{shift}}$ values obtained from Fraunhofer pattern fits in each sweep direction versus the F layer thickness. Indeed, for both NiFeCo and NiFe, $H_{\mathrm{shift}}$ vs. $d_F$ increases proportional to the magnetic flux in the junction contributed by the uniform magnetization of the ferromagnet. The trend is approximately linear due to the fact that our $\lambda_L \gg d_F $. We fit these data to:
	\begin{equation}
	\label{eqn:Hshift}
	H_{\mathrm{shift}}=M (d_F-d_\mathrm{dead})/(2 \lambda_L+ d_F+ 2 d_{\mathrm{Cu}}),
	\end{equation}
	with M and $d_\mathrm{dead}$ as free parameters. The resulting fits for NiFeCo give M = 855 $\pm$ 81 kA/m with no discernable dead layer, $d_\mathrm{dead}$ = 0.06 $\pm$ 0.17 nm, while for NiFe, M = 711 $\pm$ 143 kA/m and $d_\mathrm{dead}= -$0.05 $\pm$ 0.55 nm.
	
\begin{figure}
		\begin{center}
			\includegraphics[width=2.8 in]{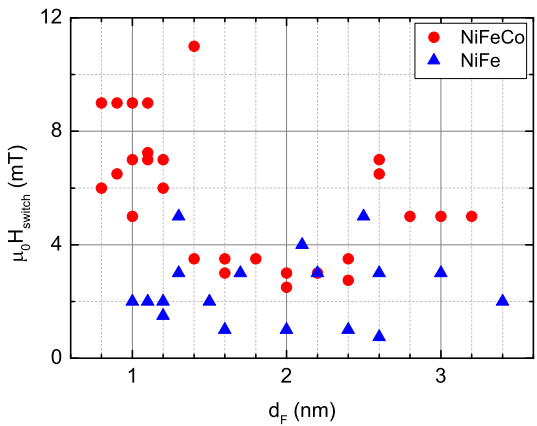}	
		\end{center}
		\caption{\label{fig:HswitchVsThickness} The average of the switching field $H_{\mathrm{switch}}$ for the two sweep directions from each Fraunhoffer pattern versus ferromagnet thickness $d_F$ for both the NiFeCo and NiFe based SFS junctions. The large increase in $H_{\mathrm{switch}}$ for NiFeCo at small values of d$_F$ is due to extrinsic factors such as strain, surface roughness, or defects.}
	\end{figure}

	In Fig.~\ref{fig:HswitchVsThickness} we plot the average of the switching field $H_{\mathrm{switch}}$ for the two sweep directions from each Fraunhofer pattern versus $d_F$. In general NiFeCo has a larger $H_{\mathrm{switch}}$ than NiFe and their difference increases as $d_F$ approaches 1 nm. Stoner-Wohlfarth theory predicts that $H_{\mathrm{switch}}$ should grow linearly with d$_F$, though that trend is clearly violated by the NiFeCo data at small d$_F$. The large scatter in the data as well as that violation, are probably the result of extrinsic factors such as surface roughness, magnetostriction, or defects in the film. The large switching field could be advantageous if NiFeCo is used as a fixed layer, since a sufficient difference between the switching fields of the free and the fixed F layers is important for controlling a cryogenic memory device.
	
\begin{figure}
		\begin{center}
			\includegraphics[width=2.8 in]{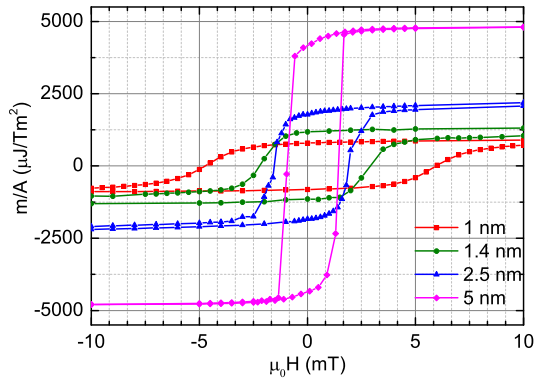}	
		\end{center}
		\caption{\label{fig:NiFeCoSQUIDMagnetomemer} NiFeCo hysteresis loops of unpatterned films with d$_{\mathrm{NiFeCo}}$ ranging from 1-5 nm. Plotted is the sample magnetic moment divided by the area versus the applied field, measured using SQUID magnetometry. Note that dividing the ordinate by the thickness in nm will give magnetization in kA/m or emu/cm$^3$. As $d_{\mathrm{NiFeCo}}$ is reduced, the switching field increases and the squareness decreases.}
	\end{figure}

	To further characterize the magnetic properties of our NiFeCo we fabricated samples for SQUID magnetometry measurements. Unpatterned thin films of Nb(25)/Cu(5)/NiFeCo(d$_F$)/Cu(5)/Au(5), with thicknesses in nanometers, were sputtered under the same conditions as for the SFS junctions. The samples were measured at 10 K in a Quantum Design SQUID magnetometer, with the applied magnetic field parallel to the plane the films. The hysteresis loops of films with $d_\mathrm{NiFeCo}$ = 1-5 nm are shown in Fig.~\ref{fig:NiFeCoSQUIDMagnetomemer}. When accounting for the F layer thickness the saturization magnetization per unit volume is nearly constant for all samples 934 $\pm$ 8 kA/m and is similar to the results in Figs.~\ref{fig:IcRn_HShift_NiFeCo}(b), while the x-intercept shows $d_{\mathrm{dead}}$= -0.06 $\pm$ 0.03 nm. Note that these unpatterned films contain many magnetic domains that switch predominantly by domain-wall motion, and should not be viewed as a direct comparison to the single-domain nanomagnets in our SFS junctions. Nonetheless, as $d_{\mathrm{NiFeCo}}$ is reduced from 5 nm to 1 nm, the coercive field increases from $\approx$ 2 mT to $\approx$ 6 mT. The Curie temperature of NiFeCo was measured to be $>$400 K, so our NiFeCo samples should not require cooling in a field to set their magnetization direction.
	
	\section{Conclusion}
	This paper provides a study of the magnetic and transport behaviors of SFS Josephson junctions containing the ferromagnetic alloys NiFeCo and NiFe. Systematic studies of the material properties of such junctions, including the 0-$\pi$ transition thickness, switching fields, and shifts in the Fraunhofer patterns versus F-layer thickness are crucial for the advancement of cryogenic memory technologies. To develop phase-controlled S/F/N/F$^\prime$/S memory devices one would need to fix either the F or F$^\prime$ layer to a thickness near it's respective 0-$\pi$ transition, while the other layer could be kept much thinner to avoid substantial decay in $I_c$. Then, by tuning the relative orientation of the F and F$^\prime$ magnetization vectors between parallel and anti-parallel configurations, the junction can be controllably toggled between the $\pi$ and 0 phase states~\cite{Gingrich2016}.
	
	We have shown that NiFeCo is a potential candidate for such a fixed layer. If positioned near the first 0-$\pi$ transition at 1.2 nm, it has a reasonable switching field, $H_{\mathrm{switch}} \approx$ 7 mT, which is somewhat larger than NiFe, but not as unwieldy as Ni (260 mT). However, from Fig.~\ref{fig:HswitchVsThickness}, it is not clear if the relatively small difference in the switching fields of NiFeCo and NiFe would allow them to be successfully used together as the fixed and free layers in a single device. One could attempt to use another material with a lower switching field than NiFe, or possibly tune the relative concentrations of Ni, Fe, and Co to optimize the switching characteristics for the situation desired.
	
	Looking forward, the addition of extra magnetic layers could pose a number of new complications: i) The surface roughness may grow with the number of layers; one would need to ensure that both the F and F$^\prime$ layers remain single domain. ii) An additional spacer layer would be required between the F and F$^\prime$ layers to keep them magnetically decoupled. As shown by Heim \textit{et al.} \cite{Heim2015}, spacer layers could cause modifications to the precise value of d$_F$ at which the 0-$\pi$ transition occurs. iii) Magnetostriction, edge roughness and other effects should be minimized so that the magnetic switching behavior of the junction is dominated primarily by shape anisotropy.
	
	Acknowledgements: We thank A. Herr, D. Miller, N. Newman, N. Rizzo and M. Eschrig for helpful discussions, B. Bi for help with fabrication using the Keck Microfabrication Facility, and A. Cramer for assistance with data analysis. This research is supported by the Office of the Director of National Intelligence (ODNI), Intelligence Advanced Research Projects Activity (IARPA), via U.S. Army Research Office contract W911NF-14-C-0115. The views and conclusions contained herein are those of the authors and should not be interpreted as necessarily representing the official policies or endorsements, either expressed or implied, of the ODNI, IARPA, or the U.S. Government. \added[id=JAG]{Sandia National Laboratories is a multimission laboratory managed and operated by National Technology and Engineering Solutions of Sandia LLC, a wholly owned subsidiary of Honeywell International Inc. for the U.S. Department of Energy’s National Nuclear Security Administration under contract DE-NA0003525.}
	\bibliography{Glick_SFS_2017}

\begin{thebibliography}{52}%
\makeatletter
\providecommand \@ifxundefined [1]{%
 \@ifx{#1\undefined}
}%
\providecommand \@ifnum [1]{%
 \ifnum #1\expandafter \@firstoftwo
 \else \expandafter \@secondoftwo
 \fi
}%
\providecommand \@ifx [1]{%
 \ifx #1\expandafter \@firstoftwo
 \else \expandafter \@secondoftwo
 \fi
}%
\providecommand \natexlab [1]{#1}%
\providecommand \enquote  [1]{``#1''}%
\providecommand \bibnamefont  [1]{#1}%
\providecommand \bibfnamefont [1]{#1}%
\providecommand \citenamefont [1]{#1}%
\providecommand \href@noop [0]{\@secondoftwo}%
\providecommand \href [0]{\begingroup \@sanitize@url \@href}%
\providecommand \@href[1]{\@@startlink{#1}\@@href}%
\providecommand \@@href[1]{\endgroup#1\@@endlink}%
\providecommand \@sanitize@url [0]{\catcode `\\12\catcode `\$12\catcode
  `\&12\catcode `\#12\catcode `\^12\catcode `\_12\catcode `\%12\relax}%
\providecommand \@@startlink[1]{}%
\providecommand \@@endlink[0]{}%
\providecommand \url  [0]{\begingroup\@sanitize@url \@url }%
\providecommand \@url [1]{\endgroup\@href {#1}{\urlprefix }}%
\providecommand \urlprefix  [0]{URL }%
\providecommand \Eprint [0]{\href }%
\providecommand \doibase [0]{http://dx.doi.org/}%
\providecommand \selectlanguage [0]{\@gobble}%
\providecommand \bibinfo  [0]{\@secondoftwo}%
\providecommand \bibfield  [0]{\@secondoftwo}%
\providecommand \translation [1]{[#1]}%
\providecommand \BibitemOpen [0]{}%
\providecommand \bibitemStop [0]{}%
\providecommand \bibitemNoStop [0]{.\EOS\space}%
\providecommand \EOS [0]{\spacefactor3000\relax}%
\providecommand \BibitemShut  [1]{\csname bibitem#1\endcsname}%
\let\auto@bib@innerbib\@empty
\bibitem [{\citenamefont {Ryazanov}\ \emph {et~al.}(2001)\citenamefont
  {Ryazanov}, \citenamefont {Oboznov}, \citenamefont {Rusanov}, \citenamefont
  {Veretennikov}, \citenamefont {Golubov},\ and\ \citenamefont
  {Aarts}}]{Ryazanov2001}%
  \BibitemOpen
  \bibfield  {author} {\bibinfo {author} {\bibfnamefont {V.~V.}\ \bibnamefont
  {Ryazanov}}, \bibinfo {author} {\bibfnamefont {V.~A.}\ \bibnamefont
  {Oboznov}}, \bibinfo {author} {\bibfnamefont {A.~Y.}\ \bibnamefont
  {Rusanov}}, \bibinfo {author} {\bibfnamefont {A.~V.}\ \bibnamefont
  {Veretennikov}}, \bibinfo {author} {\bibfnamefont {A.~A.}\ \bibnamefont
  {Golubov}}, \ and\ \bibinfo {author} {\bibfnamefont {J.}~\bibnamefont
  {Aarts}},\ }\href@noop {} {\bibfield  {journal} {\bibinfo  {journal} {Phys.
  Rev. Lett.}\ }\textbf {\bibinfo {volume} {86}},\ \bibinfo {pages} {2427}
  (\bibinfo {year} {2001})}\BibitemShut {NoStop}%
\bibitem [{\citenamefont {Kontos}\ \emph {et~al.}(2002)\citenamefont {Kontos},
  \citenamefont {Aprili}, \citenamefont {Lesueur}, \citenamefont {Gen{\^e}t},
  \citenamefont {Stephanidis},\ and\ \citenamefont {Boursier}}]{Kontos2002}%
  \BibitemOpen
  \bibfield  {author} {\bibinfo {author} {\bibfnamefont {T.}~\bibnamefont
  {Kontos}}, \bibinfo {author} {\bibfnamefont {M.}~\bibnamefont {Aprili}},
  \bibinfo {author} {\bibfnamefont {J.}~\bibnamefont {Lesueur}}, \bibinfo
  {author} {\bibfnamefont {F.}~\bibnamefont {Gen{\^e}t}}, \bibinfo {author}
  {\bibfnamefont {B.}~\bibnamefont {Stephanidis}}, \ and\ \bibinfo {author}
  {\bibfnamefont {R.}~\bibnamefont {Boursier}},\ }\href {\doibase
  10.1103/PhysRevLett.89.137007} {\bibfield  {journal} {\bibinfo  {journal}
  {Phys. Rev. Lett.}\ }\textbf {\bibinfo {volume} {89}},\ \bibinfo {pages}
  {137007} (\bibinfo {year} {2002})}\BibitemShut {NoStop}%
\bibitem [{\citenamefont {Blum}\ \emph {et~al.}(2002)\citenamefont {Blum},
  \citenamefont {Tsukernik}, \citenamefont {Karpovski},\ and\ \citenamefont
  {Palevski}}]{Blum2002}%
  \BibitemOpen
  \bibfield  {author} {\bibinfo {author} {\bibfnamefont {Y.}~\bibnamefont
  {Blum}}, \bibinfo {author} {\bibfnamefont {A.}~\bibnamefont {Tsukernik}},
  \bibinfo {author} {\bibfnamefont {M.}~\bibnamefont {Karpovski}}, \ and\
  \bibinfo {author} {\bibfnamefont {A.}~\bibnamefont {Palevski}},\ }\href
  {\doibase 10.1103/PhysRevLett.89.187004} {\bibfield  {journal} {\bibinfo
  {journal} {Phys. Rev. Lett.}\ }\textbf {\bibinfo {volume} {89}},\ \bibinfo
  {pages} {187004} (\bibinfo {year} {2002})}\BibitemShut {NoStop}%
\bibitem [{\citenamefont {Sellier}\ \emph {et~al.}(2003)\citenamefont
  {Sellier}, \citenamefont {Baraduc}, \citenamefont {Lefloch},\ and\
  \citenamefont {Calemczuk}}]{Sellier2003}%
  \BibitemOpen
  \bibfield  {author} {\bibinfo {author} {\bibfnamefont {H.}~\bibnamefont
  {Sellier}}, \bibinfo {author} {\bibfnamefont {C.}~\bibnamefont {Baraduc}},
  \bibinfo {author} {\bibfnamefont {F.}~\bibnamefont {Lefloch}}, \ and\
  \bibinfo {author} {\bibfnamefont {R.}~\bibnamefont {Calemczuk}},\ }\href
  {\doibase 10.1103/PhysRevB.68.054531} {\bibfield  {journal} {\bibinfo
  {journal} {Phys. Rev. B}\ }\textbf {\bibinfo {volume} {68}},\ \bibinfo
  {pages} {054531} (\bibinfo {year} {2003})}\BibitemShut {NoStop}%
\bibitem [{\citenamefont {Shelukhin}\ \emph {et~al.}(2006)\citenamefont
  {Shelukhin}, \citenamefont {Tsukernik}, \citenamefont {Karpovski},
  \citenamefont {Blum}, \citenamefont {Efetov}, \citenamefont {Volkov},
  \citenamefont {Champel}, \citenamefont {Eschrig}, \citenamefont
  {L{\"o}fwander}, \citenamefont {Sch{\"o}n},\ and\ \citenamefont
  {Palevski}}]{Shelukhin2006}%
  \BibitemOpen
  \bibfield  {author} {\bibinfo {author} {\bibfnamefont {V.}~\bibnamefont
  {Shelukhin}}, \bibinfo {author} {\bibfnamefont {A.}~\bibnamefont
  {Tsukernik}}, \bibinfo {author} {\bibfnamefont {M.}~\bibnamefont
  {Karpovski}}, \bibinfo {author} {\bibfnamefont {Y.}~\bibnamefont {Blum}},
  \bibinfo {author} {\bibfnamefont {K.~B.}\ \bibnamefont {Efetov}}, \bibinfo
  {author} {\bibfnamefont {A.~F.}\ \bibnamefont {Volkov}}, \bibinfo {author}
  {\bibfnamefont {T.}~\bibnamefont {Champel}}, \bibinfo {author} {\bibfnamefont
  {M.}~\bibnamefont {Eschrig}}, \bibinfo {author} {\bibfnamefont
  {T.}~\bibnamefont {L{\"o}fwander}}, \bibinfo {author} {\bibfnamefont
  {G.}~\bibnamefont {Sch{\"o}n}}, \ and\ \bibinfo {author} {\bibfnamefont
  {A.}~\bibnamefont {Palevski}},\ }\href {\doibase 10.1103/PhysRevB.73.174506}
  {\bibfield  {journal} {\bibinfo  {journal} {Phys. Rev. B}\ }\textbf {\bibinfo
  {volume} {73}},\ \bibinfo {pages} {174506} (\bibinfo {year}
  {2006})}\BibitemShut {NoStop}%
\bibitem [{\citenamefont {Weides}\ \emph {et~al.}(2006)\citenamefont {Weides},
  \citenamefont {Kemmler}, \citenamefont {Goldobin}, \citenamefont {Koelle},
  \citenamefont {Kleiner}, \citenamefont {Kohlstedt},\ and\ \citenamefont
  {Buzdin}}]{Weides2006}%
  \BibitemOpen
  \bibfield  {author} {\bibinfo {author} {\bibfnamefont {M.}~\bibnamefont
  {Weides}}, \bibinfo {author} {\bibfnamefont {M.}~\bibnamefont {Kemmler}},
  \bibinfo {author} {\bibfnamefont {E.}~\bibnamefont {Goldobin}}, \bibinfo
  {author} {\bibfnamefont {D.}~\bibnamefont {Koelle}}, \bibinfo {author}
  {\bibfnamefont {R.}~\bibnamefont {Kleiner}}, \bibinfo {author} {\bibfnamefont
  {H.}~\bibnamefont {Kohlstedt}}, \ and\ \bibinfo {author} {\bibfnamefont
  {A.}~\bibnamefont {Buzdin}},\ }\href@noop {} {\bibfield  {journal} {\bibinfo
  {journal} {Appl. Phys. Lett.}\ }\textbf {\bibinfo {volume} {89}},\ \bibinfo
  {pages} {122511} (\bibinfo {year} {2006})}\BibitemShut {NoStop}%
\bibitem [{\citenamefont {Robinson}\ \emph {et~al.}(2006)\citenamefont
  {Robinson}, \citenamefont {Piano}, \citenamefont {Burnell}, \citenamefont
  {Bell},\ and\ \citenamefont {Blamire}}]{Robinson2006}%
  \BibitemOpen
  \bibfield  {author} {\bibinfo {author} {\bibfnamefont {J.~W.~A.}\
  \bibnamefont {Robinson}}, \bibinfo {author} {\bibfnamefont {S.}~\bibnamefont
  {Piano}}, \bibinfo {author} {\bibfnamefont {G.}~\bibnamefont {Burnell}},
  \bibinfo {author} {\bibfnamefont {C.}~\bibnamefont {Bell}}, \ and\ \bibinfo
  {author} {\bibfnamefont {M.~G.}\ \bibnamefont {Blamire}},\ }\href {\doibase
  10.1103/PhysRevLett.97.177003} {\bibfield  {journal} {\bibinfo  {journal}
  {Phys. Rev. Lett.}\ }\textbf {\bibinfo {volume} {97}},\ \bibinfo {pages}
  {177003} (\bibinfo {year} {2006})}\BibitemShut {NoStop}%
\bibitem [{\citenamefont {Robinson}\ \emph {et~al.}(2007)\citenamefont
  {Robinson}, \citenamefont {Piano}, \citenamefont {Burnell}, \citenamefont
  {Bell},\ and\ \citenamefont {Blamire}}]{Robinson2007}%
  \BibitemOpen
  \bibfield  {author} {\bibinfo {author} {\bibfnamefont {J.~W.~A.}\
  \bibnamefont {Robinson}}, \bibinfo {author} {\bibfnamefont {S.}~\bibnamefont
  {Piano}}, \bibinfo {author} {\bibfnamefont {G.}~\bibnamefont {Burnell}},
  \bibinfo {author} {\bibfnamefont {C.}~\bibnamefont {Bell}}, \ and\ \bibinfo
  {author} {\bibfnamefont {M.~G.}\ \bibnamefont {Blamire}},\ }\href {\doibase
  10.1103/PhysRevB.76.094522} {\bibfield  {journal} {\bibinfo  {journal} {Phys.
  Rev. B}\ }\textbf {\bibinfo {volume} {76}},\ \bibinfo {pages} {094522}
  (\bibinfo {year} {2007})}\BibitemShut {NoStop}%
\bibitem [{\citenamefont {Bannykh}\ \emph {et~al.}(2009)\citenamefont
  {Bannykh}, \citenamefont {Pfeiffer}, \citenamefont {Stolyarov}, \citenamefont
  {Batov}, \citenamefont {Ryazanov},\ and\ \citenamefont
  {Weides}}]{Bannykh2009}%
  \BibitemOpen
  \bibfield  {author} {\bibinfo {author} {\bibfnamefont {A.~A.}\ \bibnamefont
  {Bannykh}}, \bibinfo {author} {\bibfnamefont {J.}~\bibnamefont {Pfeiffer}},
  \bibinfo {author} {\bibfnamefont {V.~S.}\ \bibnamefont {Stolyarov}}, \bibinfo
  {author} {\bibfnamefont {I.~E.}\ \bibnamefont {Batov}}, \bibinfo {author}
  {\bibfnamefont {V.~V.}\ \bibnamefont {Ryazanov}}, \ and\ \bibinfo {author}
  {\bibfnamefont {M.}~\bibnamefont {Weides}},\ }\href {\doibase
  10.1103/PhysRevB.79.054501} {\bibfield  {journal} {\bibinfo  {journal} {Phys.
  Rev. B}\ }\textbf {\bibinfo {volume} {79}},\ \bibinfo {pages} {054501}
  (\bibinfo {year} {2009})}\BibitemShut {NoStop}%
\bibitem [{\citenamefont {Terzioglu}\ and\ \citenamefont
  {Beasley}(1998)}]{Terzioglu1998}%
  \BibitemOpen
  \bibfield  {author} {\bibinfo {author} {\bibfnamefont {E.}~\bibnamefont
  {Terzioglu}}\ and\ \bibinfo {author} {\bibfnamefont {M.~R.}\ \bibnamefont
  {Beasley}},\ }\href {\doibase 10.1109/77.678441} {\bibfield  {journal}
  {\bibinfo  {journal} {IEEE Trans. Appl. Supercond.}\ }\textbf {\bibinfo
  {volume} {8}},\ \bibinfo {pages} {48} (\bibinfo {year} {1998})}\BibitemShut
  {NoStop}%
\bibitem [{\citenamefont {Ioffe}\ \emph {et~al.}(1999)\citenamefont {Ioffe},
  \citenamefont {Geshkenbein}, \citenamefont {Feigel'man}, \citenamefont
  {Fauchere},\ and\ \citenamefont {Blatter}}]{Ioffe1999}%
  \BibitemOpen
  \bibfield  {author} {\bibinfo {author} {\bibfnamefont {L.~B.}\ \bibnamefont
  {Ioffe}}, \bibinfo {author} {\bibfnamefont {V.~B.}\ \bibnamefont
  {Geshkenbein}}, \bibinfo {author} {\bibfnamefont {M.~V.}\ \bibnamefont
  {Feigel'man}}, \bibinfo {author} {\bibfnamefont {A.~L.}\ \bibnamefont
  {Fauchere}}, \ and\ \bibinfo {author} {\bibfnamefont {G.}~\bibnamefont
  {Blatter}},\ }\href {\doibase 10.1038/19464} {\bibfield  {journal} {\bibinfo
  {journal} {Nature}\ }\textbf {\bibinfo {volume} {398}},\ \bibinfo {pages}
  {679} (\bibinfo {year} {1999})}\BibitemShut {NoStop}%
\bibitem [{\citenamefont {Blatter}, \citenamefont {Geshkenbein},\ and\
  \citenamefont {Ioffe}(2001)}]{Blatter2001}%
  \BibitemOpen
  \bibfield  {author} {\bibinfo {author} {\bibfnamefont {G.}~\bibnamefont
  {Blatter}}, \bibinfo {author} {\bibfnamefont {V.~B.}\ \bibnamefont
  {Geshkenbein}}, \ and\ \bibinfo {author} {\bibfnamefont {L.~B.}\ \bibnamefont
  {Ioffe}},\ }\href {\doibase 10.1103/PhysRevB.63.174511} {\bibfield  {journal}
  {\bibinfo  {journal} {Phys. Rev. B}\ }\textbf {\bibinfo {volume} {63}},\
  \bibinfo {pages} {174511} (\bibinfo {year} {2001})}\BibitemShut {NoStop}%
\bibitem [{\citenamefont {Ustinov}\ and\ \citenamefont
  {Kaplunenko}(2003)}]{UstinovKaplunenko2003}%
  \BibitemOpen
  \bibfield  {author} {\bibinfo {author} {\bibfnamefont {A.~V.}\ \bibnamefont
  {Ustinov}}\ and\ \bibinfo {author} {\bibfnamefont {V.~K.}\ \bibnamefont
  {Kaplunenko}},\ }\href@noop {} {\bibfield  {journal} {\bibinfo  {journal} {J.
  App. Phys.}\ }\textbf {\bibinfo {volume} {94}},\ \bibinfo {pages} {5405}
  (\bibinfo {year} {2003})}\BibitemShut {NoStop}%
\bibitem [{\citenamefont {Yamashita}\ \emph {et~al.}(2005)\citenamefont
  {Yamashita}, \citenamefont {Tanikawa}, \citenamefont {Takahashi},\ and\
  \citenamefont {Maekawa}}]{Yamashita2005}%
  \BibitemOpen
  \bibfield  {author} {\bibinfo {author} {\bibfnamefont {T.}~\bibnamefont
  {Yamashita}}, \bibinfo {author} {\bibfnamefont {K.}~\bibnamefont {Tanikawa}},
  \bibinfo {author} {\bibfnamefont {S.}~\bibnamefont {Takahashi}}, \ and\
  \bibinfo {author} {\bibfnamefont {S.}~\bibnamefont {Maekawa}},\ }\href
  {\doibase 10.1103/PhysRevLett.95.097001} {\bibfield  {journal} {\bibinfo
  {journal} {Phys. Rev. Lett.}\ }\textbf {\bibinfo {volume} {95}},\ \bibinfo
  {pages} {097001} (\bibinfo {year} {2005})}\BibitemShut {NoStop}%
\bibitem [{\citenamefont {Khabipov}(2010)}]{Khabipov2010}%
  \BibitemOpen
  \bibfield  {author} {\bibinfo {author} {\bibfnamefont {M.~I.}\ \bibnamefont
  {Khabipov}},\ }\href {\doibase 10.1088/0953-2048/23/4/045032} {\bibfield
  {journal} {\bibinfo  {journal} {Supercond. Sci. Technol.}\ }\textbf {\bibinfo
  {volume} {23}},\ \bibinfo {pages} {045032} (\bibinfo {year}
  {2010})}\BibitemShut {NoStop}%
\bibitem [{\citenamefont {Feofanov}\ \emph {et~al.}(2010)\citenamefont
  {Feofanov}, \citenamefont {Oboznov}, \citenamefont {Bol'ginov}, \citenamefont
  {Lisenfeld}, \citenamefont {Poletto}, \citenamefont {Ryazanov}, \citenamefont
  {Rossolenko}, \citenamefont {Khabipov}, \citenamefont {Balashov},
  \citenamefont {Zorin}, \citenamefont {Dmitriev}, \citenamefont {Koshelets},\
  and\ \citenamefont {Ustinov}}]{Feofanov2010}%
  \BibitemOpen
  \bibfield  {author} {\bibinfo {author} {\bibfnamefont {A.~K.}\ \bibnamefont
  {Feofanov}}, \bibinfo {author} {\bibfnamefont {V.~A.}\ \bibnamefont
  {Oboznov}}, \bibinfo {author} {\bibfnamefont {V.~V.}\ \bibnamefont
  {Bol'ginov}}, \bibinfo {author} {\bibfnamefont {J.}~\bibnamefont
  {Lisenfeld}}, \bibinfo {author} {\bibfnamefont {S.}~\bibnamefont {Poletto}},
  \bibinfo {author} {\bibfnamefont {V.~V.}\ \bibnamefont {Ryazanov}}, \bibinfo
  {author} {\bibfnamefont {A.~N.}\ \bibnamefont {Rossolenko}}, \bibinfo
  {author} {\bibfnamefont {M.}~\bibnamefont {Khabipov}}, \bibinfo {author}
  {\bibfnamefont {D.}~\bibnamefont {Balashov}}, \bibinfo {author}
  {\bibfnamefont {A.~B.}\ \bibnamefont {Zorin}}, \bibinfo {author}
  {\bibfnamefont {P.~N.}\ \bibnamefont {Dmitriev}}, \bibinfo {author}
  {\bibfnamefont {V.~P.}\ \bibnamefont {Koshelets}}, \ and\ \bibinfo {author}
  {\bibfnamefont {A.~V.}\ \bibnamefont {Ustinov}},\ }\href@noop {} {\bibfield
  {journal} {\bibinfo  {journal} {Nat. Phys.}\ }\textbf {\bibinfo {volume}
  {6}},\ \bibinfo {pages} {593} (\bibinfo {year} {2010})}\BibitemShut {NoStop}%
\bibitem [{\citenamefont {Niedzielski}\ \emph {et~al.}(2015)\citenamefont
  {Niedzielski}, \citenamefont {Gingrich}, \citenamefont {Loloee},
  \citenamefont {Pratt},\ and\ \citenamefont {Birge}}]{Niedzielski2015}%
  \BibitemOpen
  \bibfield  {author} {\bibinfo {author} {\bibfnamefont {B.~M.}\ \bibnamefont
  {Niedzielski}}, \bibinfo {author} {\bibfnamefont {E.~C.}\ \bibnamefont
  {Gingrich}}, \bibinfo {author} {\bibfnamefont {R.}~\bibnamefont {Loloee}},
  \bibinfo {author} {\bibfnamefont {W.~P.}\ \bibnamefont {Pratt}}, \ and\
  \bibinfo {author} {\bibfnamefont {N.~O.}\ \bibnamefont {Birge}},\ }\href@noop
  {} {\bibfield  {journal} {\bibinfo  {journal} {Supercond. Sci. Technol.}\
  }\textbf {\bibinfo {volume} {28}},\ \bibinfo {pages} {085012} (\bibinfo
  {year} {2015})}\BibitemShut {NoStop}%
\bibitem [{\citenamefont {Gingrich}\ \emph {et~al.}(2016)\citenamefont
  {Gingrich}, \citenamefont {Niedzielski}, \citenamefont {Glick}, \citenamefont
  {Wang}, \citenamefont {Miller}, \citenamefont {Loloee}, \citenamefont {{Pratt
  Jr}},\ and\ \citenamefont {Birge}}]{Gingrich2016}%
  \BibitemOpen
  \bibfield  {author} {\bibinfo {author} {\bibfnamefont {E.~C.}\ \bibnamefont
  {Gingrich}}, \bibinfo {author} {\bibfnamefont {B.~M.}\ \bibnamefont
  {Niedzielski}}, \bibinfo {author} {\bibfnamefont {J.~A.}\ \bibnamefont
  {Glick}}, \bibinfo {author} {\bibfnamefont {Y.}~\bibnamefont {Wang}},
  \bibinfo {author} {\bibfnamefont {D.~L.}\ \bibnamefont {Miller}}, \bibinfo
  {author} {\bibfnamefont {R.}~\bibnamefont {Loloee}}, \bibinfo {author}
  {\bibfnamefont {W.~P.}\ \bibnamefont {{Pratt Jr}}}, \ and\ \bibinfo {author}
  {\bibfnamefont {N.~O.}\ \bibnamefont {Birge}},\ }\href@noop {} {\bibfield
  {journal} {\bibinfo  {journal} {Nat. Phys.}\ }\textbf {\bibinfo {volume}
  {12}},\ \bibinfo {pages} {564} (\bibinfo {year} {2016})}\BibitemShut
  {NoStop}%
\bibitem [{\citenamefont {Bell}\ \emph {et~al.}(2004)\citenamefont {Bell},
  \citenamefont {Burnell}, \citenamefont {Leung}, \citenamefont {Tarte},
  \citenamefont {Kang},\ and\ \citenamefont {Blamire}}]{Bell2004}%
  \BibitemOpen
  \bibfield  {author} {\bibinfo {author} {\bibfnamefont {C.}~\bibnamefont
  {Bell}}, \bibinfo {author} {\bibfnamefont {G.}~\bibnamefont {Burnell}},
  \bibinfo {author} {\bibfnamefont {C.~W.}\ \bibnamefont {Leung}}, \bibinfo
  {author} {\bibfnamefont {E.~J.}\ \bibnamefont {Tarte}}, \bibinfo {author}
  {\bibfnamefont {D.-J.}\ \bibnamefont {Kang}}, \ and\ \bibinfo {author}
  {\bibfnamefont {M.~G.}\ \bibnamefont {Blamire}},\ }\href@noop {} {\bibfield
  {journal} {\bibinfo  {journal} {Appl. Phys. Lett.}\ }\textbf {\bibinfo
  {volume} {84}},\ \bibinfo {pages} {1153} (\bibinfo {year}
  {2004})}\BibitemShut {NoStop}%
\bibitem [{\citenamefont {Herr}\ and\ \citenamefont
  {Herr}(2012)}]{herr_patent2012}%
  \BibitemOpen
  \bibfield  {author} {\bibinfo {author} {\bibfnamefont {A.~Y.}\ \bibnamefont
  {Herr}}\ and\ \bibinfo {author} {\bibfnamefont {Q.~P.}\ \bibnamefont
  {Herr}},\ }\href@noop {} {\enquote {\bibinfo {title} {{Josephson magnetic
  random access memory system and method}},}\ } (\bibinfo {year} {2012}),\
  \bibinfo {note} {$\mathrm{US}$ Patent 8,270,209}\BibitemShut {NoStop}%
\bibitem [{\citenamefont {Larkin}\ \emph {et~al.}(2012)\citenamefont {Larkin},
  \citenamefont {Bol’ginov}, \citenamefont {Stolyarov}, \citenamefont
  {Ryazanov}, \citenamefont {Vernik}, \citenamefont {Tolpygo},\ and\
  \citenamefont {Mukhanov}}]{Larkin2012}%
  \BibitemOpen
  \bibfield  {author} {\bibinfo {author} {\bibfnamefont {T.~I.}\ \bibnamefont
  {Larkin}}, \bibinfo {author} {\bibfnamefont {V.~V.}\ \bibnamefont
  {Bol’ginov}}, \bibinfo {author} {\bibfnamefont {V.~S.}\ \bibnamefont
  {Stolyarov}}, \bibinfo {author} {\bibfnamefont {V.~V.}\ \bibnamefont
  {Ryazanov}}, \bibinfo {author} {\bibfnamefont {I.~V.}\ \bibnamefont
  {Vernik}}, \bibinfo {author} {\bibfnamefont {S.~K.}\ \bibnamefont {Tolpygo}},
  \ and\ \bibinfo {author} {\bibfnamefont {O.~A.}\ \bibnamefont {Mukhanov}},\
  }\href@noop {} {\bibfield  {journal} {\bibinfo  {journal} {Appl. Phys.
  Lett.}\ }\textbf {\bibinfo {volume} {100}},\ \bibinfo {eid} {222601}
  (\bibinfo {year} {2012})}\BibitemShut {NoStop}%
\bibitem [{\citenamefont {Goldobin}\ \emph {et~al.}(2013)\citenamefont
  {Goldobin}, \citenamefont {Sickinger}, \citenamefont {Weides}, \citenamefont
  {Ruppelt}, \citenamefont {Kohlstedt}, \citenamefont {Kleiner},\ and\
  \citenamefont {Koelle}}]{Goldobin2013}%
  \BibitemOpen
  \bibfield  {author} {\bibinfo {author} {\bibfnamefont {E.}~\bibnamefont
  {Goldobin}}, \bibinfo {author} {\bibfnamefont {H.}~\bibnamefont {Sickinger}},
  \bibinfo {author} {\bibfnamefont {M.}~\bibnamefont {Weides}}, \bibinfo
  {author} {\bibfnamefont {N.}~\bibnamefont {Ruppelt}}, \bibinfo {author}
  {\bibfnamefont {H.}~\bibnamefont {Kohlstedt}}, \bibinfo {author}
  {\bibfnamefont {R.}~\bibnamefont {Kleiner}}, \ and\ \bibinfo {author}
  {\bibfnamefont {D.}~\bibnamefont {Koelle}},\ }\href@noop {} {\bibfield
  {journal} {\bibinfo  {journal} {App. Phys. Lett.}\ }\textbf {\bibinfo
  {volume} {102}},\ \bibinfo {eid} {242602} (\bibinfo {year}
  {2013})}\BibitemShut {NoStop}%
\bibitem [{\citenamefont {Baek}\ \emph {et~al.}(2014)\citenamefont {Baek},
  \citenamefont {Rippard}, \citenamefont {Benz}, \citenamefont {Russek},\ and\
  \citenamefont {Dresselhaus}}]{Baek2014}%
  \BibitemOpen
  \bibfield  {author} {\bibinfo {author} {\bibfnamefont {B.}~\bibnamefont
  {Baek}}, \bibinfo {author} {\bibfnamefont {W.~H.}\ \bibnamefont {Rippard}},
  \bibinfo {author} {\bibfnamefont {S.~P.}\ \bibnamefont {Benz}}, \bibinfo
  {author} {\bibfnamefont {S.~E.}\ \bibnamefont {Russek}}, \ and\ \bibinfo
  {author} {\bibfnamefont {P.~D.}\ \bibnamefont {Dresselhaus}},\ }\href
  {\doibase 10.1038/ncomms4888} {\bibfield  {journal} {\bibinfo  {journal}
  {Nature Commun.}\ }\textbf {\bibinfo {volume} {5}},\ \bibinfo {pages} {3888}
  (\bibinfo {year} {2014})}\BibitemShut {NoStop}%
\bibitem [{\citenamefont {Abd El~Qader}\ \emph {et~al.}(2014)\citenamefont {Abd
  El~Qader}, \citenamefont {Singh}, \citenamefont {Galvin}, \citenamefont {Yu},
  \citenamefont {Rowell},\ and\ \citenamefont {Newman}}]{Qader2014}%
  \BibitemOpen
  \bibfield  {author} {\bibinfo {author} {\bibfnamefont {M.}~\bibnamefont {Abd
  El~Qader}}, \bibinfo {author} {\bibfnamefont {R.~K.}\ \bibnamefont {Singh}},
  \bibinfo {author} {\bibfnamefont {S.~N.}\ \bibnamefont {Galvin}}, \bibinfo
  {author} {\bibfnamefont {L.}~\bibnamefont {Yu}}, \bibinfo {author}
  {\bibfnamefont {J.~M.}\ \bibnamefont {Rowell}}, \ and\ \bibinfo {author}
  {\bibfnamefont {N.}~\bibnamefont {Newman}},\ }\href@noop {} {\bibfield
  {journal} {\bibinfo  {journal} {Appl. Phys. Lett.}\ }\textbf {\bibinfo
  {volume} {104}},\ \bibinfo {eid} {022602} (\bibinfo {year}
  {2014})}\BibitemShut {NoStop}%
\bibitem [{\citenamefont {Herr}, \citenamefont {Herr},\ and\ \citenamefont
  {Naaman}(2015)}]{herr_phasepatent2015}%
  \BibitemOpen
  \bibfield  {author} {\bibinfo {author} {\bibfnamefont {A.}~\bibnamefont
  {Herr}}, \bibinfo {author} {\bibfnamefont {Q.}~\bibnamefont {Herr}}, \ and\
  \bibinfo {author} {\bibfnamefont {O.}~\bibnamefont {Naaman}},\ }\href@noop {}
  {\enquote {\bibinfo {title} {{Phase hysteretic magnetic Josephson junction
  memory cell}},}\ } (\bibinfo {year} {2015}),\ \bibinfo {note} {$\mathrm{US}$
  Patent App. 14/043,360}\BibitemShut {NoStop}%
\bibitem [{\citenamefont {Bakurskiy}\ \emph {et~al.}(2013)\citenamefont
  {Bakurskiy}, \citenamefont {Klenov}, \citenamefont {Soloviev}, \citenamefont
  {Bol'ginov}, \citenamefont {Ryazanov}, \citenamefont {Vernik}, \citenamefont
  {Mukhanov}, \citenamefont {Kupriyanov},\ and\ \citenamefont
  {Golubov}}]{Bakurskiy2013}%
  \BibitemOpen
  \bibfield  {author} {\bibinfo {author} {\bibfnamefont {S.~V.}\ \bibnamefont
  {Bakurskiy}}, \bibinfo {author} {\bibfnamefont {N.~V.}\ \bibnamefont
  {Klenov}}, \bibinfo {author} {\bibfnamefont {I.~I.}\ \bibnamefont
  {Soloviev}}, \bibinfo {author} {\bibfnamefont {V.~V.}\ \bibnamefont
  {Bol'ginov}}, \bibinfo {author} {\bibfnamefont {V.~V.}\ \bibnamefont
  {Ryazanov}}, \bibinfo {author} {\bibfnamefont {I.~V.}\ \bibnamefont
  {Vernik}}, \bibinfo {author} {\bibfnamefont {O.~A.}\ \bibnamefont
  {Mukhanov}}, \bibinfo {author} {\bibfnamefont {M.~Y.}\ \bibnamefont
  {Kupriyanov}}, \ and\ \bibinfo {author} {\bibfnamefont {A.~A.}\ \bibnamefont
  {Golubov}},\ }\href@noop {} {\bibfield  {journal} {\bibinfo  {journal} {App.
  Phys. Lett.}\ }\textbf {\bibinfo {volume} {102}},\ \bibinfo {eid} {192603}
  (\bibinfo {year} {2013})}\BibitemShut {NoStop}%
\bibitem [{\citenamefont {Vernik}\ \emph {et~al.}(2013)\citenamefont {Vernik},
  \citenamefont {Bol'ginov}, \citenamefont {Bakurskiy}, \citenamefont
  {Golubov}, \citenamefont {Kupriyanov}, \citenamefont {Ryazanov},\ and\
  \citenamefont {Mukhanov}}]{Vernik2013}%
  \BibitemOpen
  \bibfield  {author} {\bibinfo {author} {\bibfnamefont {I.~V.}\ \bibnamefont
  {Vernik}}, \bibinfo {author} {\bibfnamefont {V.~V.}\ \bibnamefont
  {Bol'ginov}}, \bibinfo {author} {\bibfnamefont {S.~V.}\ \bibnamefont
  {Bakurskiy}}, \bibinfo {author} {\bibfnamefont {A.~A.}\ \bibnamefont
  {Golubov}}, \bibinfo {author} {\bibfnamefont {M.~Y.}\ \bibnamefont
  {Kupriyanov}}, \bibinfo {author} {\bibfnamefont {V.~V.}\ \bibnamefont
  {Ryazanov}}, \ and\ \bibinfo {author} {\bibfnamefont {O.~A.}\ \bibnamefont
  {Mukhanov}},\ }\href@noop {} {\bibfield  {journal} {\bibinfo  {journal} {IEEE
  Transactions on Applied Superconductivity}\ }\textbf {\bibinfo {volume}
  {23}},\ \bibinfo {pages} {1701208} (\bibinfo {year} {2013})}\BibitemShut
  {NoStop}%
\bibitem [{\citenamefont {Ruppelt}\ \emph {et~al.}(2015)\citenamefont
  {Ruppelt}, \citenamefont {Sickinger}, \citenamefont {Menditto}, \citenamefont
  {Goldobin}, \citenamefont {Koelle}, \citenamefont {Kleiner}, \citenamefont
  {Vavra},\ and\ \citenamefont {Kohlstedt}}]{Ruppelt2015}%
  \BibitemOpen
  \bibfield  {author} {\bibinfo {author} {\bibfnamefont {N.}~\bibnamefont
  {Ruppelt}}, \bibinfo {author} {\bibfnamefont {H.}~\bibnamefont {Sickinger}},
  \bibinfo {author} {\bibfnamefont {R.}~\bibnamefont {Menditto}}, \bibinfo
  {author} {\bibfnamefont {E.}~\bibnamefont {Goldobin}}, \bibinfo {author}
  {\bibfnamefont {D.}~\bibnamefont {Koelle}}, \bibinfo {author} {\bibfnamefont
  {R.}~\bibnamefont {Kleiner}}, \bibinfo {author} {\bibfnamefont
  {O.}~\bibnamefont {Vavra}}, \ and\ \bibinfo {author} {\bibfnamefont
  {H.}~\bibnamefont {Kohlstedt}},\ }\href {\doibase 10.1063/1.4905672}
  {\bibfield  {journal} {\bibinfo  {journal} {Applied Physics Letters}\
  }\textbf {\bibinfo {volume} {106}},\ \bibinfo {pages} {022602} (\bibinfo
  {year} {2015})}\BibitemShut {NoStop}%
\bibitem [{\citenamefont {Bakurskiy}\ \emph {et~al.}(2017)\citenamefont
  {Bakurskiy}, \citenamefont {Filippov}, \citenamefont {Ruzhickiy},
  \citenamefont {Klenov}, \citenamefont {Soloviev}, \citenamefont
  {Kupriyanov},\ and\ \citenamefont {Golubov}}]{Bakurskiy2017}%
  \BibitemOpen
  \bibfield  {author} {\bibinfo {author} {\bibfnamefont {S.~V.}\ \bibnamefont
  {Bakurskiy}}, \bibinfo {author} {\bibfnamefont {V.~I.}\ \bibnamefont
  {Filippov}}, \bibinfo {author} {\bibfnamefont {V.~I.}\ \bibnamefont
  {Ruzhickiy}}, \bibinfo {author} {\bibfnamefont {N.~V.}\ \bibnamefont
  {Klenov}}, \bibinfo {author} {\bibfnamefont {I.~I.}\ \bibnamefont
  {Soloviev}}, \bibinfo {author} {\bibfnamefont {M.~Y.}\ \bibnamefont
  {Kupriyanov}}, \ and\ \bibinfo {author} {\bibfnamefont {A.~A.}\ \bibnamefont
  {Golubov}},\ }\href {\doibase 10.1103/PhysRevB.95.094522} {\bibfield
  {journal} {\bibinfo  {journal} {Phys. Rev. B}\ }\textbf {\bibinfo {volume}
  {95}},\ \bibinfo {pages} {094522} (\bibinfo {year} {2017})}\BibitemShut
  {NoStop}%
\bibitem [{\citenamefont {Wang}, \citenamefont {{Pratt Jr}},\ and\
  \citenamefont {Birge}(2012)}]{Wang2012}%
  \BibitemOpen
  \bibfield  {author} {\bibinfo {author} {\bibfnamefont {Y.}~\bibnamefont
  {Wang}}, \bibinfo {author} {\bibfnamefont {W.~P.}\ \bibnamefont {{Pratt
  Jr}}}, \ and\ \bibinfo {author} {\bibfnamefont {N.~O.}\ \bibnamefont
  {Birge}},\ }\href {\doibase 10.1103/PhysRevB.85.214522} {\bibfield  {journal}
  {\bibinfo  {journal} {Phys. Rev. B}\ }\textbf {\bibinfo {volume} {85}},\
  \bibinfo {pages} {214522} (\bibinfo {year} {2012})}\BibitemShut {NoStop}%
\bibitem [{\citenamefont {Thomas}, \citenamefont {Ulmer},\ and\ \citenamefont
  {Ketterson}(1998)}]{Thomas1998}%
  \BibitemOpen
  \bibfield  {author} {\bibinfo {author} {\bibfnamefont {C.~D.}\ \bibnamefont
  {Thomas}}, \bibinfo {author} {\bibfnamefont {M.~P.}\ \bibnamefont {Ulmer}}, \
  and\ \bibinfo {author} {\bibfnamefont {J.~B.}\ \bibnamefont {Ketterson}},\
  }\href@noop {} {\bibfield  {journal} {\bibinfo  {journal} {J. App. Phys.}\
  }\textbf {\bibinfo {volume} {84}},\ \bibinfo {pages} {364} (\bibinfo {year}
  {1998})}\BibitemShut {NoStop}%
\bibitem [{\citenamefont {Kohlstedt}\ \emph {et~al.}(1996)\citenamefont
  {Kohlstedt}, \citenamefont {König}, \citenamefont {Henne}, \citenamefont
  {Thyssen},\ and\ \citenamefont {Caputo}}]{Kohlstedt1996}%
  \BibitemOpen
  \bibfield  {author} {\bibinfo {author} {\bibfnamefont {H.}~\bibnamefont
  {Kohlstedt}}, \bibinfo {author} {\bibfnamefont {F.}~\bibnamefont {König}},
  \bibinfo {author} {\bibfnamefont {P.}~\bibnamefont {Henne}}, \bibinfo
  {author} {\bibfnamefont {N.}~\bibnamefont {Thyssen}}, \ and\ \bibinfo
  {author} {\bibfnamefont {P.}~\bibnamefont {Caputo}},\ }\href@noop {}
  {\bibfield  {journal} {\bibinfo  {journal} {J. App. Phys.}\ }\textbf
  {\bibinfo {volume} {80}},\ \bibinfo {pages} {5512} (\bibinfo {year}
  {1996})}\BibitemShut {NoStop}%
\bibitem [{\citenamefont {Kotula}, \citenamefont {Keenan},\ and\ \citenamefont
  {Michael}(2006)}]{Kotula2006}%
  \BibitemOpen
  \bibfield  {author} {\bibinfo {author} {\bibfnamefont {P.~G.}\ \bibnamefont
  {Kotula}}, \bibinfo {author} {\bibfnamefont {M.~R.}\ \bibnamefont {Keenan}},
  \ and\ \bibinfo {author} {\bibfnamefont {J.~R.}\ \bibnamefont {Michael}},\
  }\href@noop {} {\bibfield  {journal} {\bibinfo  {journal} {Micros. and
  Microanal.}\ }\textbf {\bibinfo {volume} {12}},\ \bibinfo {pages} {36}
  (\bibinfo {year} {2006})}\BibitemShut {NoStop}%
\bibitem [{Note1()}]{Note1}%
  \BibitemOpen
  \bibinfo {note} {Strictly speaking, the field is uniform only inside a
  uniformly magnetized ellipsoid. Because the elliptical nanomagnets in our
  junctions are very thin, there is very little difference between an ellipse
  and an ellipsoid.}\BibitemShut {Stop}%
\bibitem [{\citenamefont {Edmunds}, \citenamefont {Pratt},\ and\ \citenamefont
  {Rowlands}(1980)}]{Edmunds1980}%
  \BibitemOpen
  \bibfield  {author} {\bibinfo {author} {\bibfnamefont {D.~L.}\ \bibnamefont
  {Edmunds}}, \bibinfo {author} {\bibfnamefont {W.~P.}\ \bibnamefont {Pratt}},
  \ and\ \bibinfo {author} {\bibfnamefont {J.~A.}\ \bibnamefont {Rowlands}},\
  }\href {\doibase 10.1063/1.1136116} {\bibfield  {journal} {\bibinfo
  {journal} {Rev. Sci. Instrum.}\ }\textbf {\bibinfo {volume} {51}},\ \bibinfo
  {pages} {1516} (\bibinfo {year} {1980})}\BibitemShut {NoStop}%
\bibitem [{\citenamefont {Barone}\ and\ \citenamefont
  {Patern{\`o}}(1982)}]{Barone1982}%
  \BibitemOpen
  \bibfield  {author} {\bibinfo {author} {\bibfnamefont {A.}~\bibnamefont
  {Barone}}\ and\ \bibinfo {author} {\bibfnamefont {G.}~\bibnamefont
  {Patern{\`o}}},\ }\href@noop {} {\emph {\bibinfo {title} {{Physics and
  applications of the Josephson effect}}}}\ (\bibinfo  {publisher} {Wiley},\
  \bibinfo {year} {1982})\BibitemShut {NoStop}%
\bibitem [{\citenamefont {Vila}\ \emph {et~al.}(2000)\citenamefont {Vila},
  \citenamefont {Park}, \citenamefont {Caballero}, \citenamefont {Bozec},
  \citenamefont {Loloee}, \citenamefont {Pratt},\ and\ \citenamefont
  {Bass}}]{Vila2000}%
  \BibitemOpen
  \bibfield  {author} {\bibinfo {author} {\bibfnamefont {L.}~\bibnamefont
  {Vila}}, \bibinfo {author} {\bibfnamefont {W.}~\bibnamefont {Park}}, \bibinfo
  {author} {\bibfnamefont {J.~A.}\ \bibnamefont {Caballero}}, \bibinfo {author}
  {\bibfnamefont {D.}~\bibnamefont {Bozec}}, \bibinfo {author} {\bibfnamefont
  {R.}~\bibnamefont {Loloee}}, \bibinfo {author} {\bibfnamefont {W.~P.}\
  \bibnamefont {Pratt}}, \ and\ \bibinfo {author} {\bibfnamefont
  {J.}~\bibnamefont {Bass}},\ }\href {\doibase
  http://dx.doi.org/10.1063/1.373586} {\bibfield  {journal} {\bibinfo
  {journal} {J. App. Phys.}\ }\textbf {\bibinfo {volume} {87}},\ \bibinfo
  {pages} {8610} (\bibinfo {year} {2000})}\BibitemShut {NoStop}%
\bibitem [{\citenamefont {Ivanchenko}\ and\ \citenamefont
  {Zil'berman}(1968)}]{IvanchenkoZilberman1969}%
  \BibitemOpen
  \bibfield  {author} {\bibinfo {author} {\bibfnamefont {Y.~M.}\ \bibnamefont
  {Ivanchenko}}\ and\ \bibinfo {author} {\bibfnamefont {L.~A.}\ \bibnamefont
  {Zil'berman}},\ }\href@noop {} {\bibfield  {journal} {\bibinfo  {journal}
  {Zh. Eksp. Teor. Fiz}\ }\textbf {\bibinfo {volume} {55}},\ \bibinfo {pages}
  {2395} (\bibinfo {year} {1968})},\ \bibinfo {note} {[\textit{Soviet Physics
  J. Exp. Theor. Phys.}, \textbf{28}, 6, (1969)]}\BibitemShut {NoStop}%
\bibitem [{\citenamefont {Ambegaokar}\ and\ \citenamefont
  {Halperin}(1969)}]{AmbegaokarHalperin1969}%
  \BibitemOpen
  \bibfield  {author} {\bibinfo {author} {\bibfnamefont {V.}~\bibnamefont
  {Ambegaokar}}\ and\ \bibinfo {author} {\bibfnamefont {B.~I.}\ \bibnamefont
  {Halperin}},\ }\href@noop {} {\bibfield  {journal} {\bibinfo  {journal}
  {Phys. Rev. Lett.}\ }\textbf {\bibinfo {volume} {22}},\ \bibinfo {pages}
  {1364} (\bibinfo {year} {1969})}\BibitemShut {NoStop}%
\bibitem [{\citenamefont {Khaire}, \citenamefont {Pratt},\ and\ \citenamefont
  {Birge}(2009)}]{Khaire2009}%
  \BibitemOpen
  \bibfield  {author} {\bibinfo {author} {\bibfnamefont {T.~S.}\ \bibnamefont
  {Khaire}}, \bibinfo {author} {\bibfnamefont {W.~P.}\ \bibnamefont {Pratt}}, \
  and\ \bibinfo {author} {\bibfnamefont {N.~O.}\ \bibnamefont {Birge}},\ }\href
  {\doibase 10.1103/PhysRevB.79.094523} {\bibfield  {journal} {\bibinfo
  {journal} {Phys. Rev. B}\ }\textbf {\bibinfo {volume} {79}},\ \bibinfo
  {pages} {094523} (\bibinfo {year} {2009})}\BibitemShut {NoStop}%
\bibitem [{Note2()}]{Note2}%
  \BibitemOpen
  \bibinfo {note} {We correct a missing factor of $\mu _0$ to the corresponding
  equation in Ref.~\protect \rev@citealpnum {Khaire2009}}\BibitemShut {NoStop}%
\bibitem [{\citenamefont {Buzdin}, \citenamefont {Bulaevskii},\ and\
  \citenamefont {Panyukov}(1982)}]{Buzdin1982}%
  \BibitemOpen
  \bibfield  {author} {\bibinfo {author} {\bibfnamefont {A.~I.}\ \bibnamefont
  {Buzdin}}, \bibinfo {author} {\bibfnamefont {L.~N.}\ \bibnamefont
  {Bulaevskii}}, \ and\ \bibinfo {author} {\bibfnamefont {S.~V.}\ \bibnamefont
  {Panyukov}},\ }\href@noop {} {\bibfield  {journal} {\bibinfo  {journal}
  {Pis'ma Zh. Eksp. Teor. Fiz.}\ }\textbf {\bibinfo {volume} {35}},\ \bibinfo
  {pages} {147} (\bibinfo {year} {1982})},\ \bibinfo {note} {[\textit{J. Exp.
  Theor. Phys. Lett.}, \textbf{35}, 4, 20 (1982)]}\BibitemShut {NoStop}%
\bibitem [{\citenamefont {Buzdin}\ and\ \citenamefont
  {Kupriyanov}(1991)}]{Buzdin1991}%
  \BibitemOpen
  \bibfield  {author} {\bibinfo {author} {\bibfnamefont {A.~I.}\ \bibnamefont
  {Buzdin}}\ and\ \bibinfo {author} {\bibfnamefont {M.~Y.~a.}\ \bibnamefont
  {Kupriyanov}},\ }\href@noop {} {\bibfield  {journal} {\bibinfo  {journal}
  {Pis'ma Zh. Eksp. Teor. Fiz.}\ }\textbf {\bibinfo {volume} {53}},\ \bibinfo
  {pages} {308} (\bibinfo {year} {1991})},\ \bibinfo {note} {[\textit{J. Exp.
  Theor. Phys. Lett.}, \textbf{53}, 6, 321 (1991)]}\BibitemShut {NoStop}%
\bibitem [{\citenamefont {Bergeret}, \citenamefont {Volkov},\ and\
  \citenamefont {Efetov}(2001)}]{Bergeret2001}%
  \BibitemOpen
  \bibfield  {author} {\bibinfo {author} {\bibfnamefont {F.~S.}\ \bibnamefont
  {Bergeret}}, \bibinfo {author} {\bibfnamefont {A.~F.}\ \bibnamefont
  {Volkov}}, \ and\ \bibinfo {author} {\bibfnamefont {K.~B.}\ \bibnamefont
  {Efetov}},\ }\href {\doibase 10.1103/PhysRevB.64.134506} {\bibfield
  {journal} {\bibinfo  {journal} {Phys. Rev. B}\ }\textbf {\bibinfo {volume}
  {64}},\ \bibinfo {pages} {134506} (\bibinfo {year} {2001})}\BibitemShut
  {NoStop}%
\bibitem [{\citenamefont {Pugach}\ \emph {et~al.}(2011)\citenamefont {Pugach},
  \citenamefont {Kupriyanov}, \citenamefont {Goldobin}, \citenamefont
  {Kleiner},\ and\ \citenamefont {Koelle}}]{Pugach2011}%
  \BibitemOpen
  \bibfield  {author} {\bibinfo {author} {\bibfnamefont {N.~G.}\ \bibnamefont
  {Pugach}}, \bibinfo {author} {\bibfnamefont {M.~Y.}\ \bibnamefont
  {Kupriyanov}}, \bibinfo {author} {\bibfnamefont {E.}~\bibnamefont
  {Goldobin}}, \bibinfo {author} {\bibfnamefont {R.}~\bibnamefont {Kleiner}}, \
  and\ \bibinfo {author} {\bibfnamefont {D.}~\bibnamefont {Koelle}},\ }\href
  {\doibase 10.1103/PhysRevB.84.144513} {\bibfield  {journal} {\bibinfo
  {journal} {Phys. Rev. B}\ }\textbf {\bibinfo {volume} {84}},\ \bibinfo
  {pages} {144513} (\bibinfo {year} {2011})}\BibitemShut {NoStop}%
\bibitem [{\citenamefont {Buzdin}(2005)}]{Buzdin_SFReview2005}%
  \BibitemOpen
  \bibfield  {author} {\bibinfo {author} {\bibfnamefont {A.~I.}\ \bibnamefont
  {Buzdin}},\ }\href {\doibase 10.1103/RevModPhys.77.935} {\bibfield  {journal}
  {\bibinfo  {journal} {Rev. Mod. Phys.}\ }\textbf {\bibinfo {volume} {77}},\
  \bibinfo {pages} {935} (\bibinfo {year} {2005})}\BibitemShut {NoStop}%
\bibitem [{\citenamefont {Radovi\ifmmode~\acute{c}\else \'{c}\fi{}},
  \citenamefont {Lazarides},\ and\ \citenamefont
  {Flytzanis}(2003)}]{Radovic2003}%
  \BibitemOpen
  \bibfield  {author} {\bibinfo {author} {\bibfnamefont {Z.}~\bibnamefont
  {Radovi\ifmmode~\acute{c}\else \'{c}\fi{}}}, \bibinfo {author} {\bibfnamefont
  {N.}~\bibnamefont {Lazarides}}, \ and\ \bibinfo {author} {\bibfnamefont
  {N.}~\bibnamefont {Flytzanis}},\ }\href {\doibase 10.1103/PhysRevB.68.014501}
  {\bibfield  {journal} {\bibinfo  {journal} {Phys. Rev. B}\ }\textbf {\bibinfo
  {volume} {68}},\ \bibinfo {pages} {014501} (\bibinfo {year}
  {2003})}\BibitemShut {NoStop}%
\bibitem [{\citenamefont {Barsic}, \citenamefont {Valls},\ and\ \citenamefont
  {Halterman}(2007)}]{Barsic2007}%
  \BibitemOpen
  \bibfield  {author} {\bibinfo {author} {\bibfnamefont {P.~H.}\ \bibnamefont
  {Barsic}}, \bibinfo {author} {\bibfnamefont {O.~T.}\ \bibnamefont {Valls}}, \
  and\ \bibinfo {author} {\bibfnamefont {K.}~\bibnamefont {Halterman}},\ }\href
  {\doibase 10.1103/PhysRevB.75.104502} {\bibfield  {journal} {\bibinfo
  {journal} {Phys. Rev. B}\ }\textbf {\bibinfo {volume} {75}},\ \bibinfo
  {pages} {104502} (\bibinfo {year} {2007})}\BibitemShut {NoStop}%
\bibitem [{\citenamefont {Bass}(2016)}]{Bass2016}%
  \BibitemOpen
  \bibfield  {author} {\bibinfo {author} {\bibfnamefont {J.}~\bibnamefont
  {Bass}},\ }\href@noop {} {\bibfield  {journal} {\bibinfo  {journal} {J. Magn.
  Magn. Mater.}\ }\textbf {\bibinfo {volume} {408}},\ \bibinfo {pages} {244 }
  (\bibinfo {year} {2016})}\BibitemShut {NoStop}%
\bibitem [{\citenamefont {Faur{\'e}}\ \emph {et~al.}(2006)\citenamefont
  {Faur{\'e}}, \citenamefont {Buzdin}, \citenamefont {Golubov},\ and\
  \citenamefont {Kupriyanov}}]{Faure2006}%
  \BibitemOpen
  \bibfield  {author} {\bibinfo {author} {\bibfnamefont {M.}~\bibnamefont
  {Faur{\'e}}}, \bibinfo {author} {\bibfnamefont {A.~I.}\ \bibnamefont
  {Buzdin}}, \bibinfo {author} {\bibfnamefont {A.~A.}\ \bibnamefont {Golubov}},
  \ and\ \bibinfo {author} {\bibfnamefont {M.~Y.}\ \bibnamefont {Kupriyanov}},\
  }\href {\doibase 10.1103/PhysRevB.73.064505} {\bibfield  {journal} {\bibinfo
  {journal} {Phys. Rev. B}\ }\textbf {\bibinfo {volume} {73}},\ \bibinfo
  {pages} {064505} (\bibinfo {year} {2006})}\BibitemShut {NoStop}%
\bibitem [{\citenamefont {Bakurskiy}\ \emph {et~al.}(2015)\citenamefont
  {Bakurskiy}, \citenamefont {Golubov}, \citenamefont {Klenov}, \citenamefont
  {Kupriyanov},\ and\ \citenamefont {Soloviev}}]{Bakurskiy2015}%
  \BibitemOpen
  \bibfield  {author} {\bibinfo {author} {\bibfnamefont {S.~V.}\ \bibnamefont
  {Bakurskiy}}, \bibinfo {author} {\bibfnamefont {A.~A.}\ \bibnamefont
  {Golubov}}, \bibinfo {author} {\bibfnamefont {N.~V.}\ \bibnamefont {Klenov}},
  \bibinfo {author} {\bibfnamefont {M.~Y.}\ \bibnamefont {Kupriyanov}}, \ and\
  \bibinfo {author} {\bibfnamefont {I.~I.}\ \bibnamefont {Soloviev}},\
  }\href@noop {} {\bibfield  {journal} {\bibinfo  {journal} {Pis'ma Zh. Eksp.
  Teor. Fiz.}\ }\textbf {\bibinfo {volume} {101}},\ \bibinfo {pages} {863}
  (\bibinfo {year} {2015})},\ \bibinfo {note} {[\textit{J. Exp. Theor. Phys.
  Lett.}, \textbf{101}, 11, 765--771 (2015)]}\BibitemShut {NoStop}%
\bibitem [{\citenamefont {Heim}\ \emph {et~al.}(2015)\citenamefont {Heim},
  \citenamefont {Pugach}, \citenamefont {Kupriyanov}, \citenamefont {Goldobin},
  \citenamefont {Koelle}, \citenamefont {Kleiner}, \citenamefont {Ruppelt},
  \citenamefont {Weides},\ and\ \citenamefont {Kohlstedt}}]{Heim2015}%
  \BibitemOpen
  \bibfield  {author} {\bibinfo {author} {\bibfnamefont {D.~M.}\ \bibnamefont
  {Heim}}, \bibinfo {author} {\bibfnamefont {N.~G.}\ \bibnamefont {Pugach}},
  \bibinfo {author} {\bibfnamefont {M.~Y.}\ \bibnamefont {Kupriyanov}},
  \bibinfo {author} {\bibfnamefont {E.}~\bibnamefont {Goldobin}}, \bibinfo
  {author} {\bibfnamefont {D.}~\bibnamefont {Koelle}}, \bibinfo {author}
  {\bibfnamefont {R.}~\bibnamefont {Kleiner}}, \bibinfo {author} {\bibfnamefont
  {N.}~\bibnamefont {Ruppelt}}, \bibinfo {author} {\bibfnamefont
  {M.}~\bibnamefont {Weides}}, \ and\ \bibinfo {author} {\bibfnamefont
  {H.}~\bibnamefont {Kohlstedt}},\ }\href@noop {} {\bibfield  {journal}
  {\bibinfo  {journal} {New J. Phys.}\ }\textbf {\bibinfo {volume} {17}},\
  \bibinfo {pages} {113022} (\bibinfo {year} {2015})}\BibitemShut {NoStop}%
\end{thebibliography}%
\end{document}